\documentclass[a4paper,fleqn,usenatbib]{mnras}

\usepackage[T1]{fontenc}
\usepackage{ae,aecompl}

\usepackage{graphicx}	% Including figure files
\usepackage{amsmath}	% Advanced maths commands
\usepackage{amssymb}	% Extra maths symbols
\usepackage{multirow}

\newcommand{\dd}{\mathrm{d}}
\newcommand{\be}{\begin{equation}}
\newcommand{\ee}{\end{equation}}
\newcommand{\ba}{\begin{eqnarray}}
\newcommand{\ea}{\end{eqnarray}}

\title[On the ISW-cluster cross-correlation in future surveys]
{On the ISW-cluster cross-correlation in future surveys}

\author[M. Ballardini, D. Paoletti, F. Finelli, L. Moscardini, B. Sartoris, L. Valenziano]{Mario Ballardini$^{1,2,3,4}$
\thanks{Contact e-mail: \href{mailto:mario.ballardini@gmail.com}{mario.ballardini@gmail.com}},
Daniela Paoletti$^{3,4}$
\thanks{Contact e-mail: \href{mailto:daniela.paoletti@inaf.it}{daniela.paoletti@inaf.it}}, 
Fabio Finelli$^{3,4}$, 
Lauro Moscardini$^{2,5,4}$,
\newauthor Barbara Sartoris$^{6,7}$, Luca Valenziano$^{3,4}$
\\
$^{1}$Department of Physics and Astronomy, University of the Western Cape, Cape Town 7535, South Africa \\
$^{2}$Dipartimento di Fisica e Astronomia, Alma Mater Studiorum
Universit\`a di Bologna, Via Gobetti, 93/2, I-40129 Bologna, Italy\\
$^{3}$INAF/OAS Bologna, via Gobetti 101, I-40129 Bologna, Italy\\
$^{4}$INFN, Sezione di Bologna, Via Berti Pichat 6/2, I-40127 Bologna, Italy\\
$^{5}$INAF/OAS Bologna, via Gobetti 93/3, I-40129 Bologna, Italy\\
$^{6}$Dipartimento di Fisica, Sezione di Astronomia, Universit\`a di Trieste,
Via Tiepolo 11, I-34143 Trieste, Italy\\
$^{7}$INAF - Osservatorio Astronomico di Trieste,
Via Tiepolo 11, I-34143 Trieste, Italy}

\date{Accepted XXX. Received YYY; in original form ZZZ}

\pubyear{2018}

\begin{document}
\label{firstpage}
\pagerange{\pageref{firstpage}--\pageref{lastpage}}
\maketitle

\begin{abstract}
We investigate the cosmological information contained in the cross-correlation between the Integrated 
Sachs-Wolfe (ISW) of the Cosmic Microwave Background (CMB) anisotropy pattern and galaxy clusters 
from future wide surveys. 
Future surveys will provide cluster catalogues with a number of objects comparable with galaxy 
catalogues currently used for the detection of the ISW signal by cross-correlation with the CMB 
anisotropy pattern. By computing the angular power spectra of clusters and the corresponding 
cross-correlation with CMB, we perform a signal-to-noise ratio (SNR) analysis for the ISW detection 
as expected from the eROSITA and the Euclid space missions. 
We discuss the dependence of the SNR of the ISW-cluster cross-correlation on the specifications of 
the catalogues and on the reference cosmology. 
We forecast that the SNRs for ISW-cluster cross-correlation are slightly smaller compared to those 
which can be obtained from future galaxy surveys but the signal is expected to be detected at high 
significance, i.e. more than $> 3\,\sigma$. We also forecast the joint constraints on parameters of 
model extensions of the concordance $\Lambda$CDM cosmology by combining CMB and the ISW-cluster 
cross-correlation.
\end{abstract}


%%%%%%%%%%%%%%%%%%%%%%%%%%%%%%%%%%%%%%%%%%%%%%%%%%

%%%%%%%%%%%%%%%%% BODY OF PAPER %%%%%%%%%%%%%%%%%%

\section{Introduction}
\label{sec:zero}
One of the key predictions of the recent accelerated expansion is the late Integrated Sachs-Wolfe 
(ISW) effect \citep{Sachs:1967er} in the Cosmic Microwave Background (CMB) anisotropy pattern. 
The ISW effect is caused by the time evolution of the gravitational potentials encountered by CMB 
photons during their journey from the last scattering surface to the observer. 
The change of the gravitational potentials on large scales is caused by the accelerated expansion of 
the Universe driven by dark energy \citep{Kofman:1985fp} and is therefore correlated with density 
fluctuations at low redshifts.

The late ISW effect is just a small fraction of the CMB temperature anisotropy signal for a 
$\Lambda$CDM concordance cosmology and is maximum at largest angular scales \citep{Kofman:1985fp}, 
where cosmic variance is dominant. 
An extraction of the ISW effect was predicted as feasible by cross-correlating CMB with large scale 
structure (LSS) tracers of the matter distribution \citep{Crittenden:1995ak}. Several measurements of 
the ISW-LSS cross-correlation have been reported since the release of the WMAP first-year data 
\citep{Nolta:2003uy} by using different dark matter (DM) tracers. These measurements have been 
performed by using different estimators in different domains (e.g. 
\cite{Fosalba:2003iy,Boughn:2003yz,Afshordi:2003xu,Vielva:2004zg,McEwen:2006my,Ho:2008bz,
Giannantonio:2008zi,HernandezMonteagudo:2009fb,Schiavon:2012fc,Bianchini:2016czu,Stolzner:2017ged}). 
See \cite{Dupe:2010zs} and references therein for a review of pre-$Planck$ ISW measurements.

A detection at 2.9$\,\sigma$ level of the ISW effect has beeen obtained by \cite{Ade:2015dva} 
by cross-correlating the $Planck$ temperature map with a compilation of publicly available galaxy 
surveys (see \cite{Ade:2013dsi} for the results based on the nominal mission). 
This cross-correlation measurement of the ISW effect is compatible with the predictions of a 
concordance $\Lambda$CDM model with $\Omega_\Lambda \sim 0.7$ \citep{Ade:2015dva}. 
Cross-correlation between CMB lensing and matter tracers improve the significance 
of the ISW detection, see \cite{Ferraro:2014msa,Manzotti:2014kta,Ade:2015dva,Shajib:2016bes}.
Note also that the ISW could be targeted by using its cross-correlation with CIB \citep{Ilic:2011hh}
or tSZ \citep{Taburet:2010hb}.

Future LSS experiments, such as DESI 
\footnote{\href{http://desi.lbl.gov/}{http://desi.lbl.gov/}} 
\citep{Levi:2013gra,Aghamousa:2016sne,Aghamousa:2016zmz},
Euclid \footnote{\href{http://sci.esa.int/euclid/}{http://sci.esa.int/euclid/}}
\citep{Laureijs:2011gra,Amendola:2012ys},
LSST \footnote{\href{http://www.lsst.org/}{http://www.lsst.org/}} \citep{Abell:2009aa},
SKA \footnote{\href{http://www.skatelescope.org/}{http://www.skatelescope.org/}}
\citep{Maartens:2015mra} will provide surveys with
a larger number of objects and sky fraction probing a larger volume, then allowing a better 
determination of the ISW signal. The next generation of LSS surveys will not only lead to an 
improvement in the ISW measurement with matter tracers already employed for this purpose, but could 
also provide different catalogues of matter tracers suitable for novel detections of the ISW effect 
\citep{Raccanelli:2014kga,Raccanelli:2015lca,Pourtsidou:2016dzn}.

Current cluster surveys are not optimal for the detection of the ISW effect since 
photometric catalogues are limited to small redshift $z\lesssim 0.4$ \citep{Rozo:2009jj} and X-ray 
catalogues cover small patches of the sky \citep{Mantz:2015vua}. 
\footnote{The only analysis of the late ISW using clusters has been performed by \cite{Granett:2008ju} 
stacking WMAP data behind 50 superclusters (and 50 supervoids) to study the residual signal from 
the detection of the thermal SZ effect.}

Otherwise, the cluster catalogues from 
the next generation of LSS surveys are indeed expected to contain a 
higher number of clusters over a wider redshift range. These future catalogues will allow precision 
cosmology with clusters, as from the cosmological parameter forecasts from the Euclid cluster survey 
\citep{Sartoris:2015aga}. It is therefore interesting to study the potential for an ISW detection by 
a cross-correlation with future cluster catalogues as matter tracers. 

In this paper, by computing the angular power spectra of cluster counts and of their cross-correlation 
with CMB, we perform a signal-to-noise ratio (SNR) analysis for the ISW detection with two 
representative examples for future cluster catalogues, as those expected from the eROSITA 
\citep{2010SPIE.7732E..0UP} and Euclid \citep{Laureijs:2011gra} space missions and we forecast 
how ISW-cluster cross-correlation could improve the constraints on cosmological parameters for 
different extended models. 
This analysis is complementary to the stacking of CMB data in correspondence of 
superclusters \citep{Granett:2008ju,Papai:2010gd,Ilic:2011hh,Ade:2013dsi,Ade:2015dva}, 
whose theoretical interpretation is more difficult, although consistent with $\Lambda$CDM.

Our paper is organized as follows. In Sec.~\ref{sec:two} we review the angular power spectrum 
of the cross-correlation between CMB and a generic matter tracer, such as clusters or galaxies. 
In Sec.~\ref{sec:three} we review the calculations of the bias and the number density distribution 
for cluster catalogues. In Sec.~\ref{sec:four} we describe the future surveys considered in our 
work: eROSITA and Euclid for clusters, Euclid photometric and spectroscopic surveys for 
galaxies.
In Sec.~\ref{sec:five} we compute the auto- and CMB cross-correlation angular power spectra for 
these surveys and perform a SNR analysis. In Sec.~\ref{sec:six} we present forecasted contraints 
on cosmological parameters of extensions of the concordance $\Lambda$CDM model 
by combining CMB and the CMB-LSS cross-correlation. We draw our conclusions in Sec.~\ref{sec:concl}.

Throughout this work we adopt a fiducial cosmological model compatible with the most recent $Planck$ 
data \citep{Aghanim:2016yuo}, corresponding to $\omega_\textrm{b}\equiv\Omega_\textrm{b}h^2=0.02214$, 
$\omega_\textrm{c}\equiv\Omega_\textrm{c}h^2=0.1206$, $H_0=66.89$, $\tau=0.0581$, 
$n_\textrm{s}=0.9625$, and $\log\left(10^{10}\ A_\textrm{s}\right)=3.053$.

\section{Cross-correlation of ISW effect with number counts}
\label{sec:two}
We consider the projected density contrast of a tracer of matter X (clusters or galaxies in 
this paper) in the direction $\hat n$ as:
\be
\delta_\mathrm{X} (\hat{n}) = \int \dd z\, b_\mathrm{X} (z) \frac{\dd N_\mathrm{X}}{\dd z} \delta_\mathrm{m} (\hat{n}, z) \,.
\label{eqn:tracer}
\ee
In Eq.~\eqref{eqn:tracer} we denote by $b_\mathrm{X}(z)$ the linear bias and by $\dd N_\mathrm{X}/\dd z$ 
the redshift distribution of the tracer.
The observed tracer density is correlated with the ISW contribution to temperature fluctuations in direction 
$\hat{m}$:
\be
\frac{\delta T_\mathrm{ISW}}{T} (\hat{m}) = - \int \dd z\, e^{-\tau (z)} \left(  \frac{\dd \Phi}{\dd z} (\hat{m}\,, z) + \frac{\dd \Psi}{\dd z} (\hat{m}\,, z)  \right) \,,
\ee
where $\Phi$ and $\Psi$ are the gravitational potentials in the longitudinal gauge and $e^{-\tau (z)}$ 
is the visibility function.
 
With these assumptions, the angular power spectrum of the tracer $C_{\ell}^\mathrm{XX}$ 
and of its cross-correlation with CMB temperature $C_{\ell}^\mathrm{TX}$ are respectively:
\be
\label{eqn:APScross}
C_{\ell}^\mathrm{XX} = 4 \pi \int \frac{\dd k}{k} \Delta^2 (k) \left[I_\ell^\mathrm{X} (k)\right]^2 \,,
\ee
\be
C_{\ell}^\mathrm{TX} = 4 \pi \int \frac{\dd k}{k} \Delta^2 (k) I_\ell^\mathrm{ISW} (k) I_\ell^\mathrm{X} (k) \,, 
\ee
where we denote by $\Delta^2 (k)$ the scale invariant matter power spectrum 
$\Delta^2 (k) \equiv k^3 P(k)/ (4\pi^2)$, and define the kernels of the counts distribution and of 
the cross-correlation as: 
\be
I_\ell^\mathrm{X} (k) = \int \dd z \, b_\mathrm{X}(z) \, \frac{\dd N_\mathrm{X}}{\dd z} D(z) j_\ell \left( k \chi(z) \right) \,,
\ee
\be
I_\ell^\mathrm{ISW} (k) = - \int \dd z \, e^{-\tau} \, \left( \frac{\dd \Phi}{\dd z} + \frac{\dd \Psi}{\dd z} \right) j_\ell \left( k \chi(z) \right) \,,
\ee
where $D(z)$ is the growth factor and $\chi(z)$ is the conformal distance.

\section{Cluster number density}
\label{sec:three}
We now review the computation of the number density for clusters and the corresponding bias 
that we will use in Eq.~\ref{eqn:APScross}.
The number density of expected clusters within the solid angle $\Delta\Omega$ is:
\be
\label{eqn:count}
\frac{\dd N(z)}{\dd z} = \int_{\Delta\Omega}\textrm{d}\Omega \,
\frac{\textrm{d}V_\textrm{c}}{\textrm{d}z\,\textrm{d}\Omega} \int_0^{\infty} \textrm{d}M \, X(M,z) n_\textrm{h}(M,z) \,,
\ee
where $V_\textrm{c}$ is the comoving volume, $n_\textrm{h}(M,z)$ and $X(M,z)$ are the dark matter 
halo mass function and the survey characteristic function, respectively.

\subsection{Halo mass function}
The distribution $n_\textrm{h}(M_\textrm{h},z)$ of DM halos in mass, position, and redshift describes 
the overall abundance and clustering properties of galaxies, and the dependence on the underlying 
cosmological model. As halo mass function, we use \citep{Tinker:2008ff}:
\begin{align}
\label{eqn:hamomassfunction}
n_\textrm{h}(M,z) &= -f(\sigma) \frac{\rho_\textrm{m}}{M} \frac{\textrm{d}\ln \sigma(M,z)}{{\textrm d} M} \,,\\
f(\sigma) &= C_1(z)\left[1+\left(\frac{\sigma(M,z)}{C_3(z)}\right)^{-C_2(z)}\right] 
e^{-C_0/\sigma(M,z)^2} \,,
\end{align}
where $\rho_\textrm{m}$ is the total physical matter density and $\sigma(M,z)$ is the root mean 
square density fluctuation within a sphere. The variance of the fractional density fluctuation:
\be
\sigma^2(M,z) = \int_0^{\infty} \frac{k^2\textrm{d}k}{2\pi^2} P(k,z) W^2(k,M) \,,
\ee
is calculated by integrating the linear matter power spectrum $P(k,z)$ smoothed by a 
top-hat window in real space window function $W(k,M)$.

According to \cite{Tinker:2008ff}, we parametrize the redshift evolution of the parameters of the 
mass function in Eq.~\eqref{eqn:hamomassfunction} as different power law functions of $(1+z)$:
\begin{align}
C_1(z) &= C_{1,0}(1+z)^{-0.14} \,,\notag\\
C_2(z) &= C_{2,0}(1+z)^{-0.06} \,,\notag\\
C_2(z) &= C_{3,0}(1+z)^{-\gamma} \,,\notag\\
\log \left(\gamma\right) &= -\left(\frac{0.75}{\log\left(\Delta_\textrm{c}/75\right)}\right)^{1.2} \,,
\end{align}
where the subscript $0$ denotes the value of the quantity at $z=0$ and $\Delta_\textrm{c}$ defines 
the mean overdensity of the spherical halos in terms of the critical density of the universe. 
The quantities $C_0, C_{1,0}, C_{2,0}, C_{3,0}$ depend 
on $\Delta_\textrm{c}$ and we refer the reader to \cite{Tinker:2008ff} for their numerical values.
We use for our forecasts $\Delta_\textrm{c} = 200$ and 500 for Euclid and eROSITA respectively.

\subsection{Halo bias function}
We now specify the halo bias function, $b_\textrm{eff}(z)$. The linear bias weighted by the halo 
mass function is:
\be
b_\textrm{eff}(z) = \frac{1}{n_\textrm{h}(z)}
\int_0^\infty \textrm{d}M n_\textrm{h}(M,z)b_\textrm{h}(M,z) X(M,z) \,,
\ee
where the integrated halo redshift distribution is:
\be
n_\textrm{h}(z) = \int_0^\infty \textrm{d}M n_\textrm{h}(M,z) X(M,z) \,.
\ee

According to \cite{Tinker:2010my}, the Eulerian halo bias is:
\be
\label{eqn:bias}
b_\textrm{h}(M_\textrm{h},z) = 1-c_1\frac{\nu^{c_4}}{\nu^{c_4}+\delta_\textrm{c}^{c_4}}+c_2\nu^{c_5}+c_3\nu^{c_6} \,,
\ee
where $\nu=\delta_\textrm{c}/\sigma(M_\textrm{h},z)$ and $\delta_\textrm{c}$ is the critical density 
for collapse (in all calculations we use $\delta_\textrm{c}=1.686$) and the parameters of the halo 
bias function in Eq.~\eqref{eqn:bias} as a function of $\Delta_\textrm{c}$ are:
\begin{align}
c_1 &= 1+0.24xe^{-\left(4/x\right)^4} \,,\notag\\
c_2 &= 0.183 \,,\notag\\
c_3 &= 0.019+0.107x+0.19e^{-\left(4/x\right)^4} \,,\notag\\
c_4 &= 0.44x-0.88 \,,\notag\\
c_5 &= 1.5 \,,\notag\\
c_6 &= 2.4 \,,\notag
\end{align}
where $x\equiv\log \Delta_\textrm{c}$.

\subsection{Survey characteristic function}
\label{sec:nuisances}
We express the survey characteristic completeness function in Eq.~\eqref{eqn:count} as:
\be
X(M,z) = \int_{M_\textrm{lim}(z)}^\infty \dd M^{\textrm{obs}}P(M^{\textrm{obs}}|M) \,,
\ee
where $M_\textrm{lim}(z)$ represents the minimum value of the observed mass for a cluster to be 
included in the survey, and it is determined by the survey selection function.
The function $P(M^{\textrm{obs}}|M)$ gives the probability that a cluster of true mass $M$ has 
a measured mass $M^\textrm{obs}$ and takes into account the uncertainties that a scaling relation 
introduces in the knowledge of the cluster mass.
Following \cite{Lima:2005tt} and \cite{Sartoris:2010cr}, the probability of assigning to a cluster 
of true mass $M$ an observed mass $M^\textrm{obs}$ can be well approximated with a %Gaussian
lognormal density distribution:
\be
P(M^{\textrm{obs}}|M) = \frac{1}{M^\textrm{obs}\sqrt{2\pi \sigma^2_{\log M}}}
\text{exp}\left[-\frac{(\ln M^\textrm{obs}-B_M-\ln M)^2}{2\sigma^2_{\log M}}\right] \,,
\ee
where the parameter $B_\mathrm{M}=B_\mathrm{M,0}+\alpha\log\left(1+z\right)$ represents the 
fractional value of the systematic bias in the mass estimate and 
$\sigma_{\log M}^2 = \sigma_{\log M,0}^2 -  1 + (1+z)^{2\beta}$ is the intrinsic scattering. 
We treat $B_{\mathrm{M},0} = 0$, $\alpha=0$, $\sigma_{\log M,0} = 0.2$, and $\beta=0.125$ as 
fiducial as in \cite{Sartoris:2015aga} and we consider them as nuisance parameters in the 
following analysis.

\section{Future Surveys}
\label{sec:four}
In this section we describe two future cluster catalogues suitable for the ISW detection by 
cross-correlation with CMB, i.e. those expected from eROSITA \citep{2010SPIE.7732E..0UP} and Euclid 
\citep{Laureijs:2011gra}. For comparison, we also present the specifications for the photometric 
and spectroscopic galaxy surveys expected from Euclid, which are expected to improve on the current 
statistical significance of the ISW detection with galaxy surveys.
\begin{figure}
\centering
\includegraphics[width=7cm]{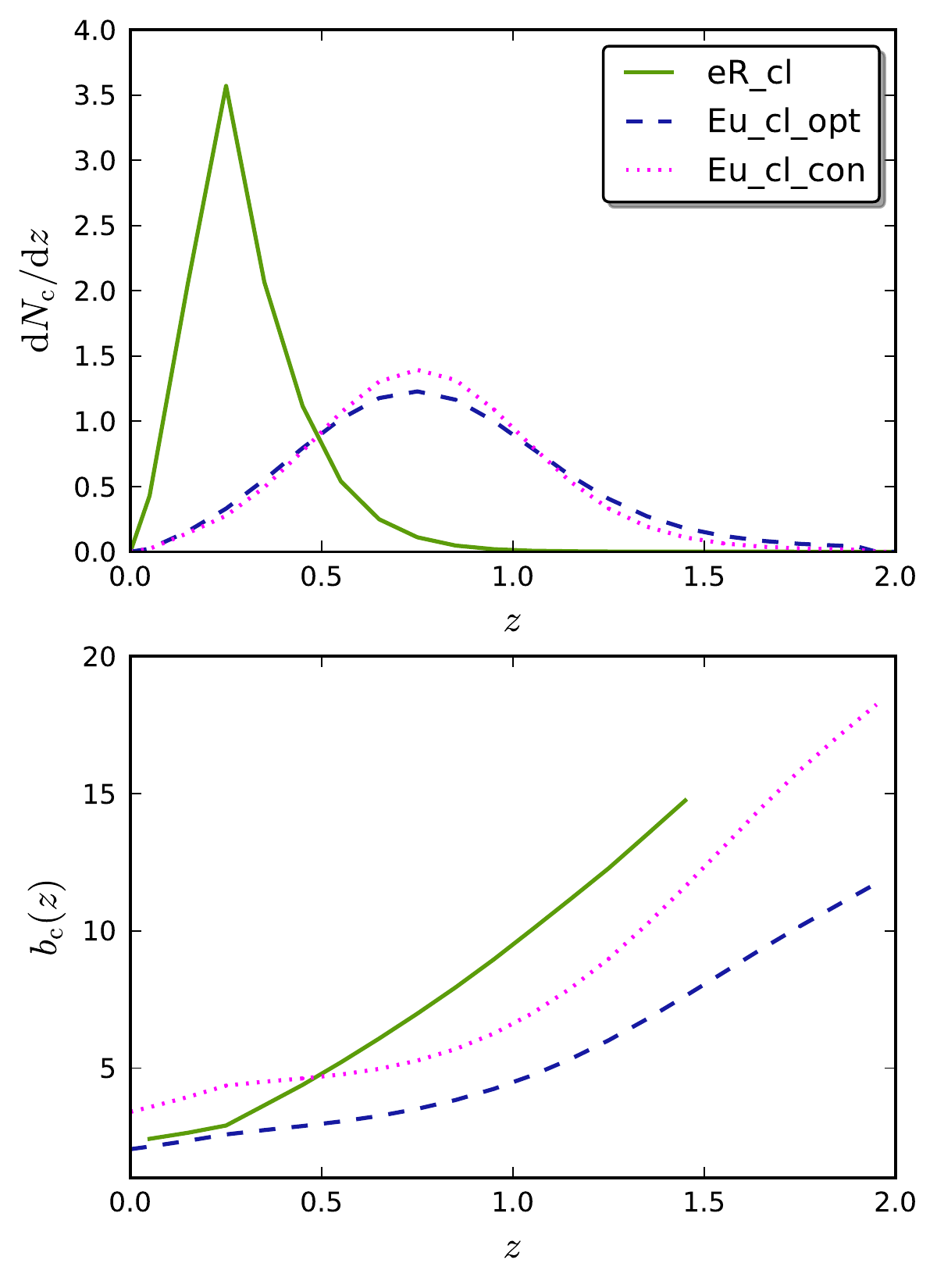}
\caption{In the top panel, we plot the normalized cluster redshift distributions adopted for each 
survey: eROSITA (solid green), optimistic Euclid (dashed blue), and conservative Euclid (dotted magenta). 
In the bottom panel, we plot the corresponding linear bias.}
\label{fig:cluster_spec}
\end{figure}

\subsection{eROSITA cluster survey}
eROSITA \footnote{Extended ROentgen Survey with an Imaging Telescope Array, 
http://www.mpe.mpg.de/erosita/} \citep{2010SPIE.7732E..0UP} is the primary science instrument onboard 
the Spectrum Roentgen-Gamma (SRG) satellite \footnote{http://hea.iki.rssi.ru/SRG/en/index.php}, 
launched in 2017.

eROSITA is expected to perform an X-ray all-sky survey with a sensitivity $\sim 30$ times better 
than ROSAT and to cover a redshift range of $0\le z \le 1.5$ with a sky coverage of 27,000 deg$^2$.
We consider the eROSITA cluster selection function presented in \cite{Pillepich:2011zz}, computed 
for a mass at $\Delta_\textrm{c}=500$ with an exposure time of $1.6\times10^3$ s and a detection 
threshold of 50 photons. In addition, we adopt as lower cut for the minimum mass 
$5\times10^{13}\ M_\odot /h$.

The resulting specifications of the eROSITA cluster catalogue (hereafter referred eR\_cl) are 
summarized in Tab.~\ref{tab:Spec}, leading to a total number of clusters $\sim 8.1\times 10^4$. 
The corresponding redshift density distribution and the linear bias for eROSITA are shown in 
Fig.~\ref{fig:cluster_spec}.

\subsection{Euclid cluster survey}
The European Space Agency (ESA) Cosmic Vision mission Euclid \citep{Laureijs:2011gra} is 
scheduled to be launched in 2021, with the goal of exploring the dark sector of the Universe. 

As selection function for the expected Euclid cluster catalogue we consider the one presented 
in \cite{Sartoris:2015aga} computed for a mass at $\Delta_\textrm{c}=200$ 
with a threshold for the 
significance of the clusters detection of 3 in terms of the ratio between the cluster galaxy 
number counts and the field rms
%$N_{500,\textrm{c}}/\sigma_\textrm{field}=3$ 
(hereafter referred Eu\_cl\_opt) with a sky coverage 
of 15,000 deg$^2$ over a redshift range of $0.2\le z \le 2$.
We also consider a detection threshold of 5
%$N_{500,\textrm{c}}/\sigma_\textrm{field}=5$ 
as a more conservative case (hereafter 
referred Eu\_cl\_con). This selection funtion was obtained following a phenomenological approach. 
In summary, the number of cluster galaxies is obtained by integration of observed cluster luminosity 
function down to the H-band magnitude limit of the Euclid survey, and the variance in the field counts 
takes into account both Poisson noise and cosmic variance (see \cite{Sartoris:2015aga} for 
more details).
Even if the galaxy cluster mass selection function used for the Euclid photometric
survey is not constant (see \cite{Sartoris:2015aga}), a mass limit of $8\times10^{13}\ M_\odot /h$ 
and $3\times10^{14}\ M_\odot /h$ approximately mimics the counts of Eu\_cl\_opt and of Eu\_cl\_con, 
respectively.

The resulting specifications of the optimistic and conservative Euclid expected cluster catalogues 
are summarized in Tab.~\ref{tab:Spec}, leading to a total number of clusters $1.3\times10^{6}$ and 
$2.0\times10^{5}$, respectively. See also Fig.~\ref{fig:cluster_spec} for the corresponding redshift 
density distribution and the linear bias.

\subsection{Euclid galaxy surveys}
\begin{figure}
\centering
\includegraphics[width=7cm]{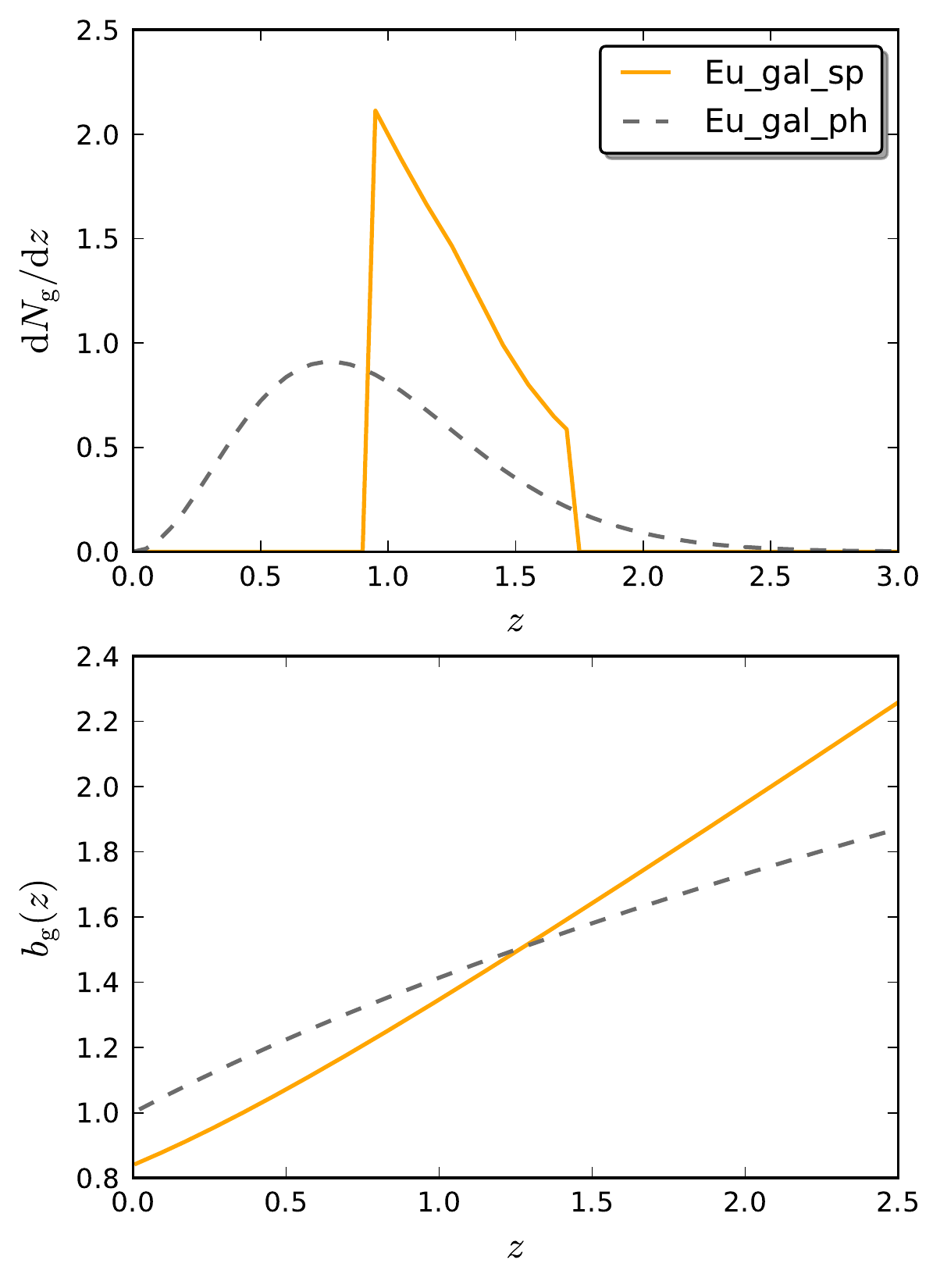}
\caption{In the top panel, we plot the normalized galaxy redshift distribution adopted for 
each survey: spectroscopic Euclid (solid orange), and photometric Euclid (dashed gray). In the 
bottom panel, we plot the corresponding linear bias.}
\label{fig:galaxy_spec}
\end{figure}

Euclid will measure the galaxy clustering in a spectroscopic survey of tens of millions of H$\alpha$ 
emitting galaxies and the cosmic shear in a photometric survey of billions of galaxies. 

For the Euclid wide spectroscopic survey (hereafter referred Eu\_gal\_sp), we consider an area of 
15,000 deg$^2$. According to the updated predictions obtained by \cite{Pozzetti:2016cch}, the Euclid 
wide single-grism survey will reach a flux limit of 
$F_{\textup{H}\alpha} > 2 \times 10^{-16}\ \text{erg}\ \text{cm}^{-2}\ \text{s}^{-1}$ and will 
cover a redshift range $0.9 \le z \le 1.8$. 

As specifications for the Euclid photometric survey (hereafter referred Eu\_gal\_ph), we adopt 
$\dd N/\dd z \propto z^2 \mathrm{exp}[-(z/z_0)^{3/2}]$, with $z_0 = z_{\mathrm mean}/1.412$ 
the peak of the distribution and $z_{\mathrm mean}$ the median \citep{Amendola:2012ys}. 
We choose $z_{\mathrm mean}=0.9$, a surface density $\bar{n}_\mathrm{g}=30$ per arcmin$^2$ and 
a bias $b_\textrm{g}(z)=\sqrt{1+z}$ \citep{Amendola:2012ys}.\\

The specifications for the Euclid spectroscopic and photometric galaxy surveys are summarized in 
Tab.~\ref{tab:Spec}.
In Fig.~\ref{fig:galaxy_spec} we plot the normalized redshift density distribution and the effective bias.

\begin{table}
\centering
\caption{
Summary of the characteristics of the surveys studied inthis paper: fraction of the sky available 
f$_\textrm{sky}$ and surface density of tracer $\bar{n}_\textrm{X}$.}
\label{tab:Spec}
\begin{tabular}{|l|c|c|}
\hline
\rule[-1mm]{0mm}{.4cm}
& f$_\textrm{sky}$ [deg$^2$] & $\bar{n}_\textrm{X}$ [deg$^{-2}$] \\
\hline
\rule[-1mm]{0mm}{.4cm}
eR\_cl & $27,000$ & $2.9$ \\
\rule[-1mm]{0mm}{.4cm}
Eu\_cl\_opt & $15,000$ & $72$ \\
\rule[-1mm]{0mm}{.4cm}
Eu\_cl\_con & $15,000$ & $10.8$ \\
\hline
\rule[-1mm]{0mm}{.4cm}
Eu\_gal\_sp & $15,000$ & $3,960$ \\
\rule[-1mm]{0mm}{.4cm}
Eu\_gal\_ph & $15,000$ & $108,000$ \\
\hline
\end{tabular}
\end{table}

\section{A signal-to-noise analysis}
\label{sec:five}
\begin{figure}
\centering
\includegraphics[width=7cm]{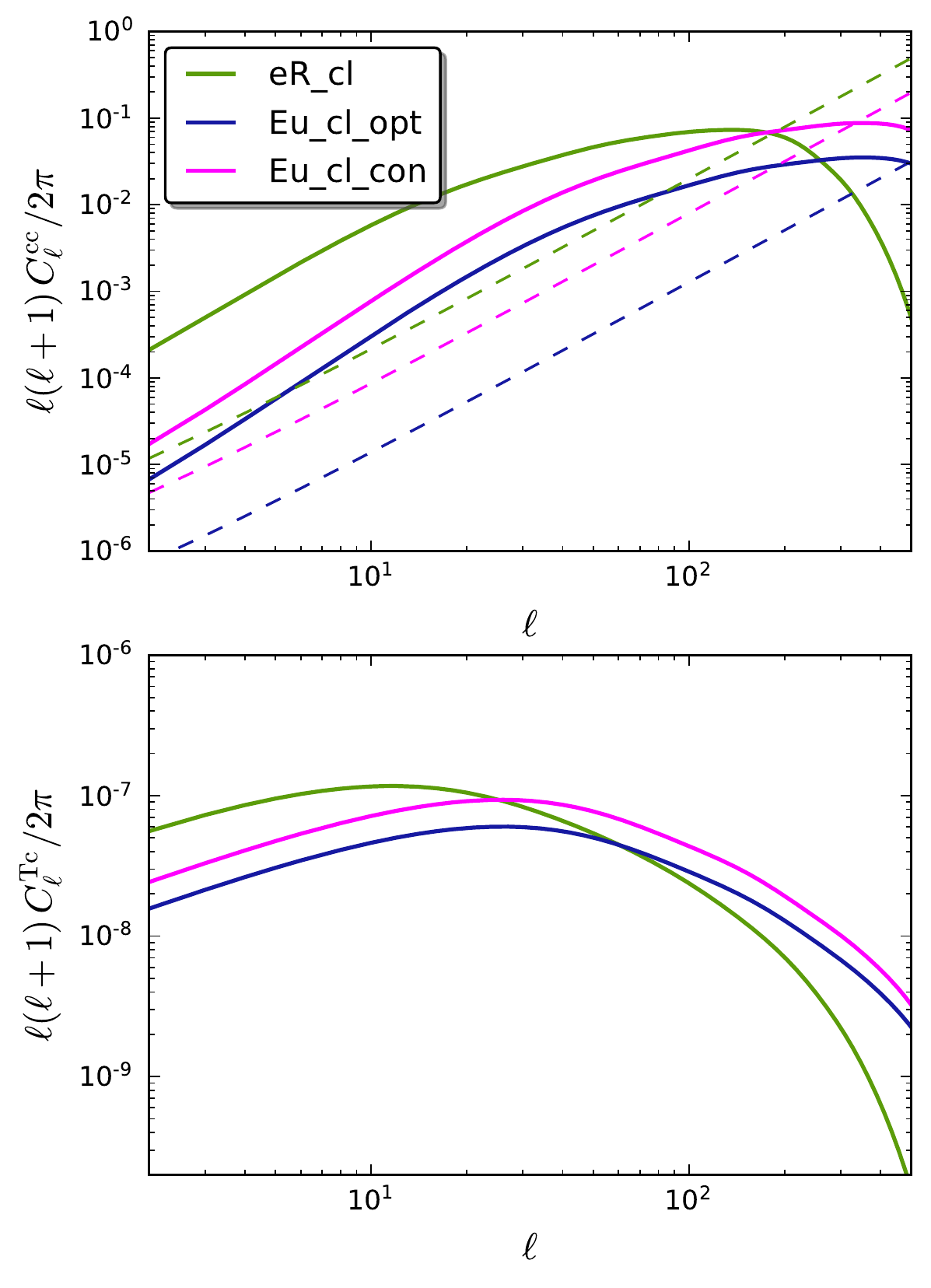}
\caption{Cluster angular power spectra $C_\ell^\textrm{cc}$ (top panel) and 
cross-correlation angular power spectra $C_\ell^\textrm{Tc}$ 
(bottom panel) for the three surveys of clusters: eROSITA (solid green), optimistic Euclid (solid 
blue), and conservative Euclid (solid magenta). In the top panel the corresponding shot-noise 
(dashed lines) is also shown.}
\label{fig:cluster_cl}
\end{figure}

\begin{figure}
\centering
\includegraphics[width=7cm]{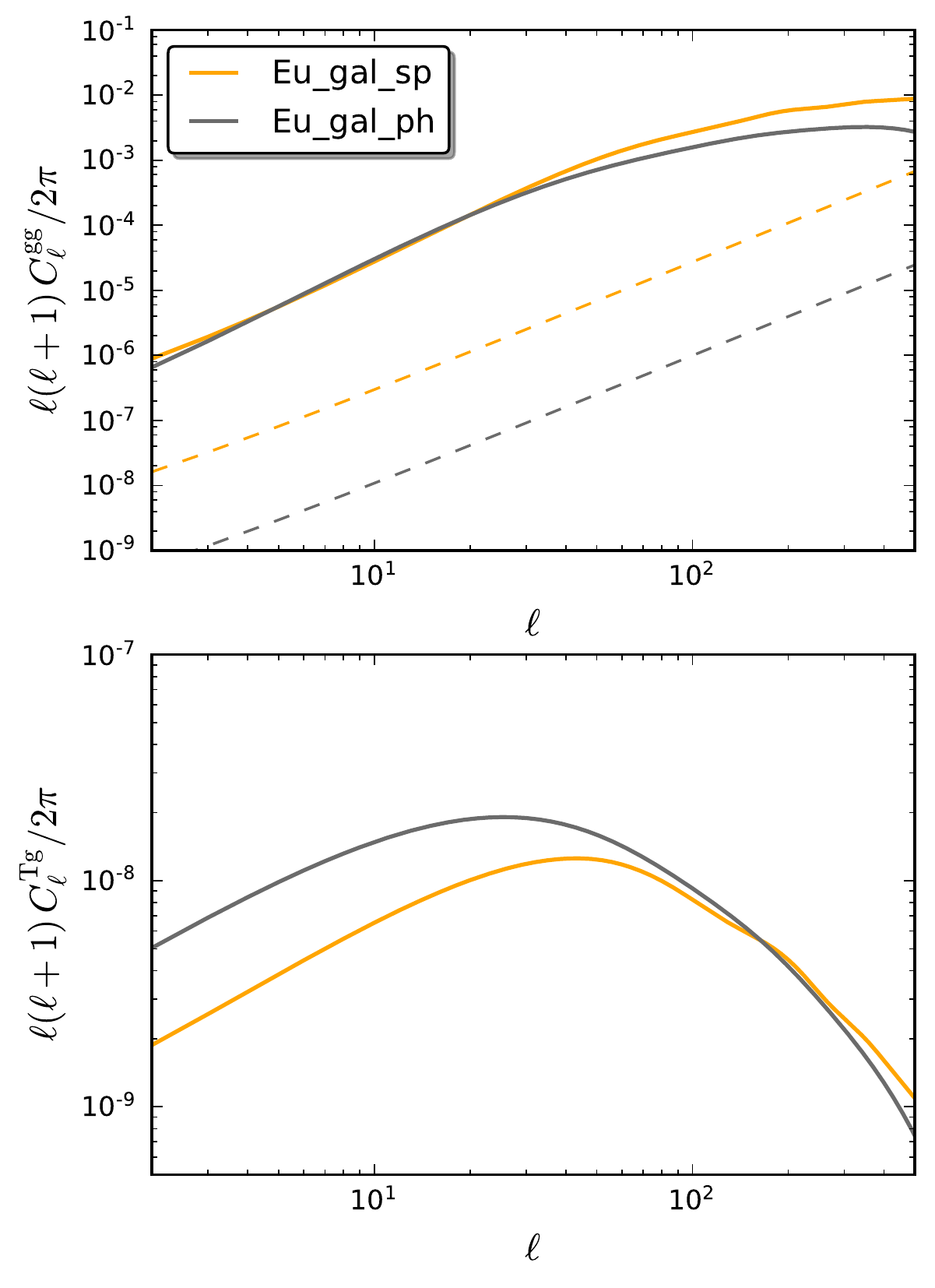}
\caption{Galaxy angular power spectra $C_\ell^\textrm{gg}$ (top panel) and 
cross-correlation angular power spectra $C_\ell^\textrm{Tg}$ 
(bottom panel) for two surveys of galaxies: spectroscopic Euclid (solid orange), and photometric 
Euclid (solid gray). In the top panel the corresponding shot-noise (dashed lines) is also shown.}
\label{fig:galaxy_cl}
\end{figure}

We now investigate the detection level of the ISW effect by computing the signal-to-noise ratio 
(SNR), which we define as \citep{Cooray:2001ab,Afshordi:2004kz}:
\be
\label{eqn:SNT}
\left(\frac{S}{N}\right)^2_\textrm{X}
= \sum_{\ell=2}^{\ell_\textrm{max}} \left(\frac{S}{N}\right)^2_{\textrm{X},\,\ell}  \,,
\ee
and
\be
\left(\frac{S}{N}\right)^2_{\textrm{T},\ell} = \left(2\ell+1\right)
\frac{f_\textrm{sky}^\textrm{X}\left(C_\ell^\mathrm{TX}\right)^2}{\left(C_\ell^\mathrm{TX}\right)^2+\bar{C}_\ell^\mathrm{TT} \bar{C}_\ell^\mathrm{XX}} \,, 
\ee
where $f_\textrm{sky}$ is the sky coverage of the survey and an overall bar stands for the sum of the 
signal and its noise. By $\bar{C}_\ell^\mathrm{TT}$ we denote the total CMB temperature angular power 
spectrum, which includes both the ISW and the Sachs-Wolfe terms, plus the effective noise:
\be
\bar{C}_\ell^\mathrm{TT} = C_\ell^\mathrm{TT} + {\cal N}_\ell^\mathrm{T} \,.
\ee
At low and intermediate multipoles, $Planck$ provides a measurement with a high 
signal-to-noise of the total anisotropy signal and the effective noise in temperature can therefore 
be neglected. The noise contribution for the mass tracer associated to a finite number of counts 
corresponds to a shot-noise as:
\be
{\cal N}_\ell^\textrm{X} = \frac{1}{\bar{n}_\mathrm{X}} \,,
\ee
where $\bar{n}_\mathrm{X}$ is the surface density of tracers per steradian.\\
We show in Fig.~\ref{fig:cluster_cl} and Fig.~\ref{fig:galaxy_cl} the angular power spectra and 
the noises, for cluster and galaxy surveys respectively, calculated with the assumed fiducial cosmology 
and used in the SNR analysis.

The ISW effect leaves also an imprint on the E-mode polarization that can be quantified in analogy 
with Eq. (\ref{eqn:SNT}), by substituting T with E. We have checked that the SNR is at most $0.2$ for 
the surveys considered here. 
Whereas the cross-correlation between the CMB E-mode polarization and LSS tracers is small as 
expected \citep{Cooray:2005yj}, the inclusion of polarization data allows a reduction of the error 
bars in the overall detection of the ISW effect \citep{2009MNRAS.395.1837F}. Indeed, the E-mode 
polarization is correlated with primary temperature anisotropy \citep{Zaldarriaga:1996ke} and 
therefore indirectly helps in disentangling the ISW effect. It is possible to extend the SNR analysis 
in Eq.~\eqref{eqn:SNT} to the inclusion of polarization as in \cite{2009MNRAS.395.1837F}:
\be
\label{eqn:SNTE}
\left(\frac{S}{N}\right)^2_{\textrm{T,\,E},\,\ell}
=\left(2\ell+1\right)
\frac{f_\textrm{sky}^\textrm{X}\left(C_\ell^\mathrm{TX}\right)^2}
{\left(C_\ell^\mathrm{TX}\right)^2+\left(\bar{C}_\ell^\mathrm{TT}-\frac{\left(C_\ell^\mathrm{TE}\right)^2}{\bar{C}_\ell^\mathrm{EE}}\right)\bar{C}_\ell^\mathrm{XX}} \,.
\ee
In the denominator of the latter equation the power spectrum of temperature anisotropies is 
substituted by the corresponding one of the E-mode uncorrelated temperature map:
\be
a_\mathrm{\ell m}^\mathrm{T} \to a_\mathrm{\ell m}^\mathrm{T} - \frac{C_\ell^\mathrm{TE}}{\bar{C}_\ell^\mathrm{EE}}a_\mathrm{\ell m}^\mathrm{E} \,.
\ee
The E-uncorrelated temperature map has been already introduced in the ISW-LSS cross-correlation 
analysis of the $Planck$ 2015 release \citep{Ade:2015dva}.

\begin{figure}
\centering
\includegraphics[width=7cm]{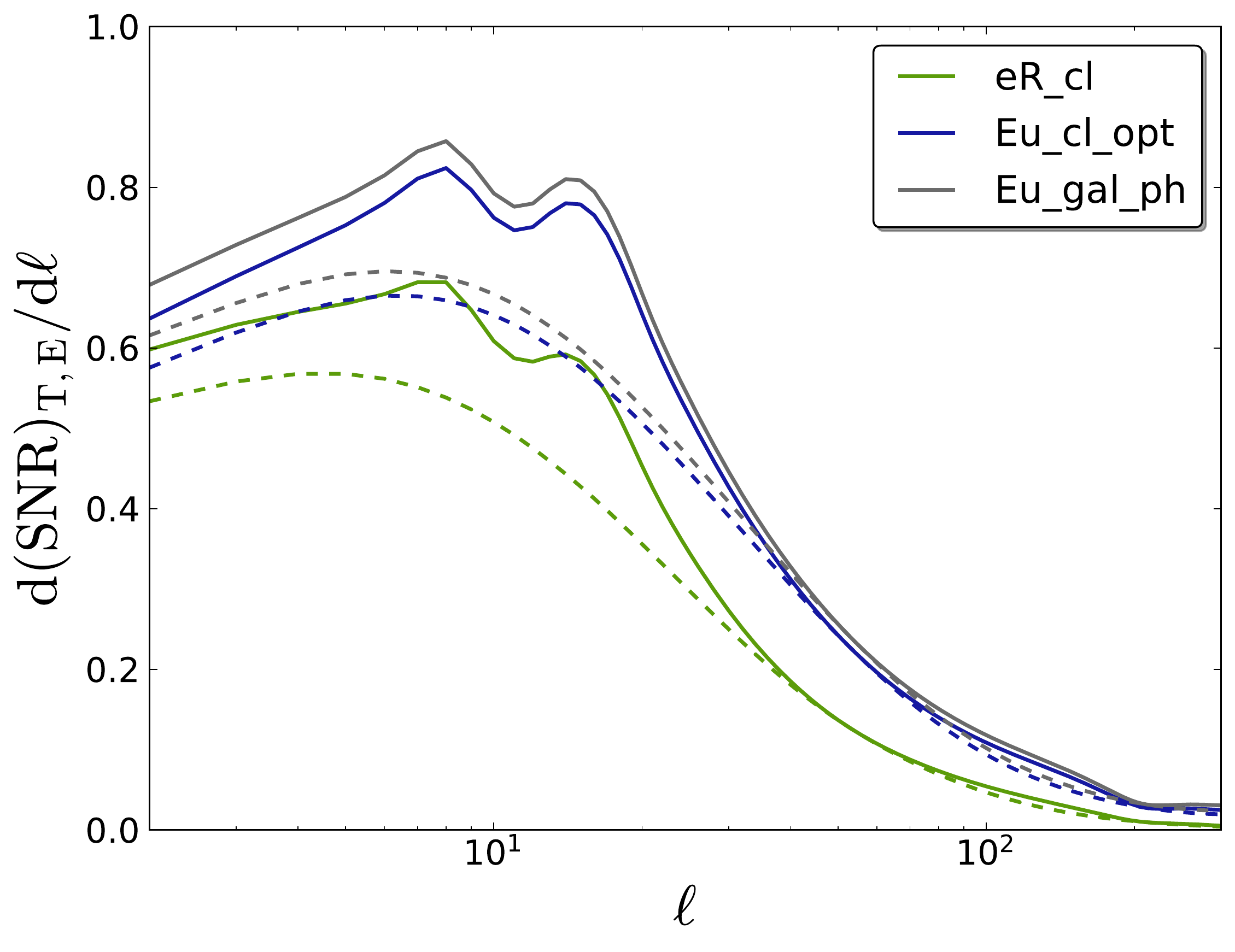}
\caption{Comparison of $(S/N)_{\textrm{X},\,\ell}$ for temperature only (dashed) and including 
polarization (solid) according to Eq.~\eqref{eqn:SNT} and Eq.~\eqref{eqn:SNTE} respectively, for 
the eROSITA and optimistic Euclid survey of clusters, together with the Euclid photometric survey 
of galaxies.}
\label{fig:SN}
\end{figure}

We summarize in Tab.~\ref{tab:SN} the SNR for the cluster and galaxy catalogues introduced in 
Sec.~\ref{sec:two} with the estimators in Eqs. (\ref{eqn:SNT},\ref{eqn:SNTE}). 
Future cluster catalogues from eROSITA and Euclid could lead to a detection of the ISW effect at high 
significance $> 3\,\sigma$ thanks to the high number of clusters detected, i.e. a small noise level. We 
find that the SNR for these two cluster surveys will be comparable to the one obtainable from future 
galaxy surveys. In particular, for a Euclid survey, the SNR from clusters is smaller than the one 
expected from the photometric survey, but it is still larger than the SNR forecasted for the 
spectroscopic survey. The reason resides in the high bias expected for clusters which compensate 
the larger shot noise in the observed cluster maps. 
The inclusion of CMB polarization information in the analysis increases the SNR by approximately 
18\,\% as compared to the temperature only, with a weak dependence of the type of survey considered; 
this is fully consistent with \cite{2009MNRAS.395.1837F,Giannantonio:2012aa}. We find the possibility 
of detecting the ISW effect at $4\,\sigma$ both the cluster and the galaxy Euclid photometric 
surveys by adding the CMB polarization information.
A comparison between the two SNR calculated from the two estimators is also shown in Fig.~\ref{fig:SN}.

\begin{table*}
\centering
\caption{SNR for ISW detection with future cluster and galaxy catalogues. The first row refers to the 
SNR with CMB temperature in Eq.~\eqref{eqn:SNT}; the second one to the estimate taking into account both CMB 
temperature and E-mode polarization as in Eq.~\eqref{eqn:SNTE}.}
\label{tab:SN}
\begin{tabular}{c|ccc|cc}
\hline
\rule[-1mm]{0mm}{.4cm}
 & eR\_cl & Eu\_cl\_opt & Eu\_cl\_con & Eu\_gal\_sp & Eu\_gal\_ph \\
\hline
\rule[-1mm]{0mm}{.4cm}
T & $2.3$ & $3.5$ & $3.3$ & $2.2$ & $3.7$ \\
\rule[-1mm]{0mm}{.4cm}
T,E & $2.8$ & $4.1$ & $3.9$ & $2.5$ & $4.3$ \\
\hline
\end{tabular}
\end{table*}

\subsection{Robustness from the specifications of the survey and cosmology}

We now test the dependence of the SNR obtainable for clusters on some of the specifications of the 
surveys, namely the smallest multipoles $\ell_\textrm{min}$, the covered sky fraction 
$f_\textrm{sky}^\textrm{X}$, the maximum redshift $z_\textrm{max}$, and the minimum detectable mass 
$M_\textrm{min}$; for the sake of brevity, we present this analysis for the optimistic Euclid cluster 
(Eu\_cl\_opt) survey only and considering the inclusion of polarization in the SNR estimation.
\begin{figure*}
\centering
\includegraphics[width=15cm]{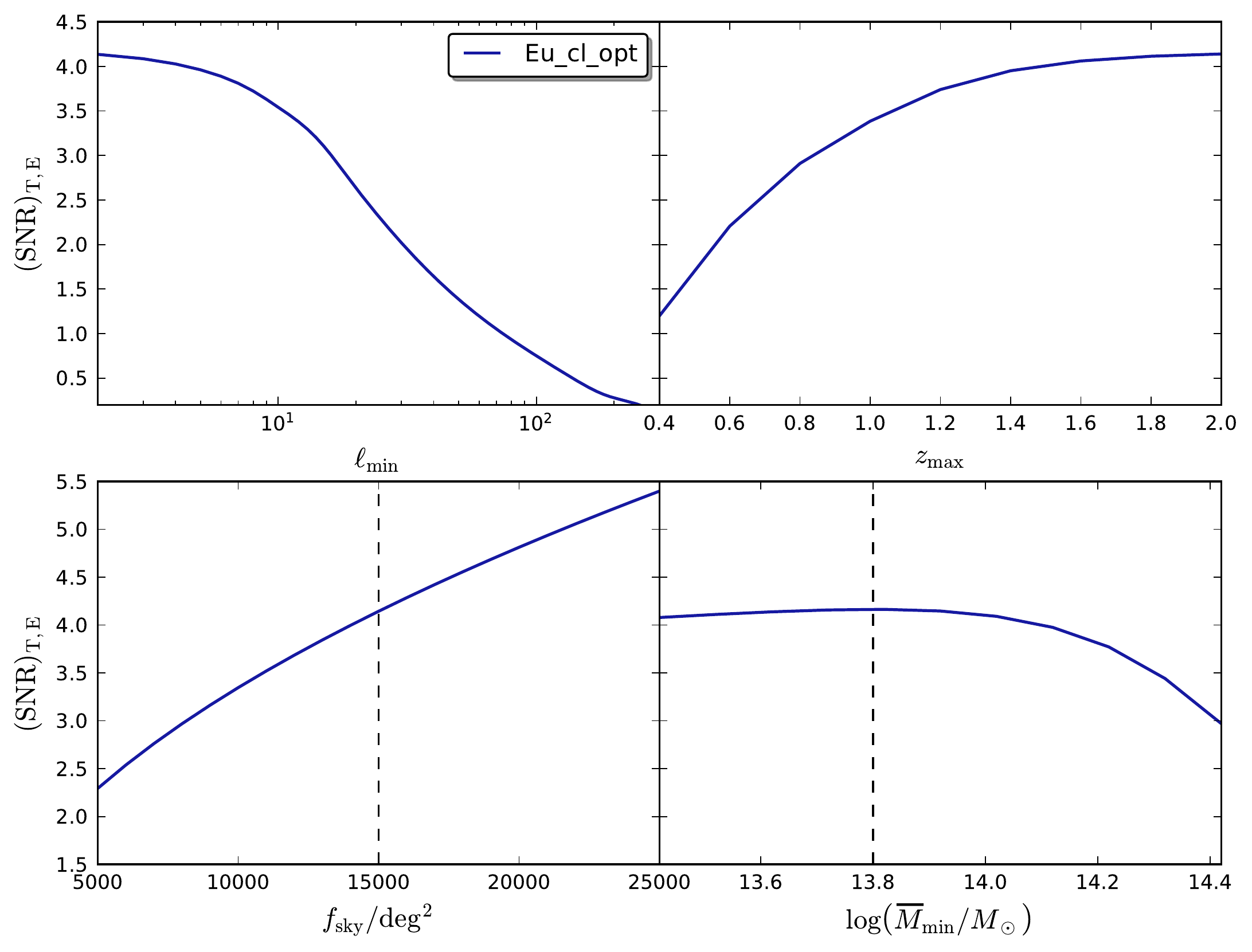}
\caption{We plot the SNR for the optimistic Euclid cluster survey (Eu\_cl\_opt) as a function of 
$\ell_\textrm{min}$ (top left), the maximum redshift probed by the survey (top right), the area of 
the sky survey (bottom left), and the minimum observed mass as over the redshift range $0.2\le z\le 2$ 
(bottom right). Dashed line represents the expected SNR for the Eu\_cl\_opt specifications. All the SNR 
take into account of the contribution from CMB E-mode polarization according to Eq.~\eqref{eqn:SNTE}.}
\label{fig:SNR}
\end{figure*}

Given the possibility to lose the first multipoles because of some mask effects and possible 
systematics on the largest scales, we show in Fig.~\ref{fig:SNR} (top left panel) the dependence of 
SNR on $\ell_\mathrm{min}$. The SNR remains approximately constant when removing the first few 
multipoles, and is decreased approximately by $0.5\,\sigma$ cutting the first ten multipoles.

Bottom left panel of Fig.~\ref{fig:SNR} shows the SNR as a function of the observed sky fraction, 
which does not affect just Eq.~\eqref{eqn:SNTE} as SNR$\ \propto \sqrt{f_\textrm{sky}^\textrm{X}}$ 
but it changes also the amount of total clusters observed, i.e. the shot-noise error.

The SNR is sufficiently robust to a reduction of $z_\textup{max}$; the forecast SNR remains above a 
$4\,\sigma$ detection for $z_\textup{max} \gtrsim 1.4$ for the optimistic Euclid cluster survey when 
the CMB polarization is added.

We finally study the dependence of the SNR when changing $M_\textrm{min}$. 
We vertically shift a redshift independent selection function around $8\times10^{13}$ M$_\odot$, 
as representative of the selection function from \citep{Sartoris:2015aga} for the photometric Euclid 
cluster survey. The results in Fig.~\ref{fig:SNR} (bottom left panel) show that the expected selection 
function for Euclid would be optimal in terms of SNR.

We can conclude that for the ISW detection is more importantan to have wide surveys, 
covering a large fraction of the sky, than deeper ones, since the surveys analysed 
are already signal dominated on the scales relevant for the ISW.

\begin{figure}
\centering
\includegraphics[width=7cm]{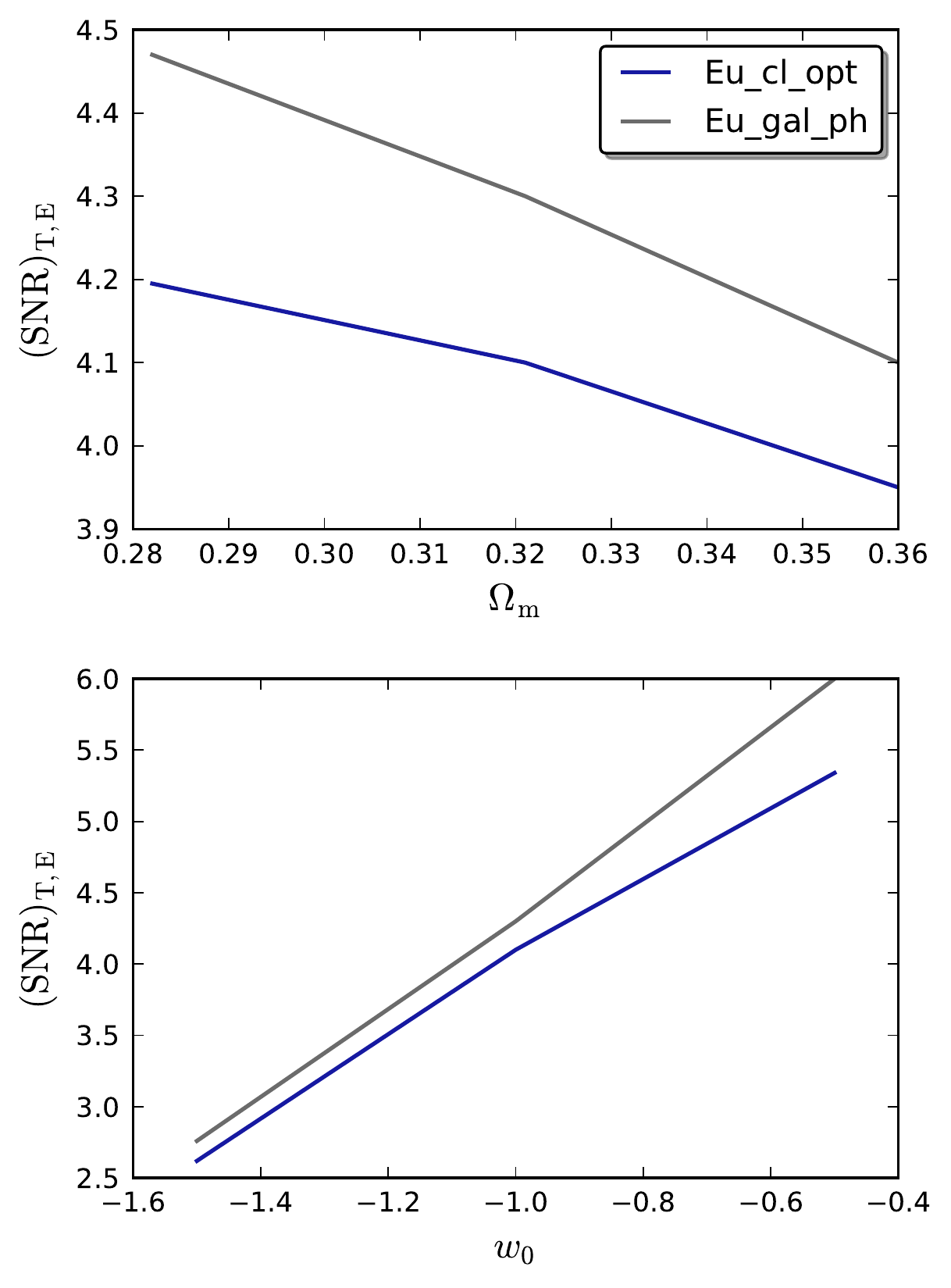}
\caption{Top panel: dependence on $\Omega_\textrm{m}$ of the SNR for the ISW detection with clusters 
(optimistic Euclid cluster survey, blue line) and with galaxies (photometric Euclid galaxy survey, 
gray line). Bottom panel: dependence on the dark energy equation of state $w_0$ of the SNR for the 
ISW detection.}
\label{fig:SNparameter}
\end{figure}

We now briefly discuss the dependence of the SNR on the underlying cosmology.
The cross-correlation power spectrum is sensitive to the total matter density $\Omega_\textrm{m}$ 
and to the dark energy equation of state $w_0$. They both change the amplitude of $C_\ell^\mathrm{TX}$ 
(since for a flat universe an increased amount of matter density corresponds to have a smaller amount 
of dark energy density). Moreover, the peak of $C_\ell^\mathrm{TX}$ is shifted according to a 
different matter-dark energy equivalence. We show the dependence of the SNR on these two parameters 
in Fig.~\ref{fig:SNparameter} by keeping fixed all the other cosmological parameters (note that we 
change $\sigma_8$ according to the variation of $\Omega_\textrm{m}$ and $w_0$).
The SNR varies less than $0.2\,\sigma$ when we change $\Omega_\textrm{m}$ up to $3\,\sigma$ away 
from the $Planck$ best-fit \citep{Aghanim:2016yuo}. We find a stronger effect on the dark energy 
parameter of state $w_0$. Even a small shift of $\Delta w_0\simeq 0.2$ would lead to a difference 
larger than $0.5\,\sigma$ on the SNR. In particular, by assuming a fiducial cosmology with a smaller 
value of $w_0$ leads to a higher SNR.

\subsection{Combining clusters with galaxies}
Different LSS surveys can be combined to increase the significance of the ISW detection. 
The best improvement comes for uncorrelated LSS surveys for which the SNR squared can be 
simply added. In this case we would obtain for the ISW detection a SNR of 3.2/4.1/4.0 
for the combination of Eu\_gal\_sp and 4.3/5.1/5.0 for the combination of Eu\_gal\_ph with 
eR\_cl, Eu\_cl\_opt, Eu\_cl\_con respectively.
However, this ideal case is usually reached for surveys which cover different redshifts 
or for distributions of tracers which peak at different redshifts.

The Euclid's cluster catalogue will be extracted from the photometric galaxy survey. 
Thus, they will trace by construction the same cosmological information mostly, 
having a similar redshift distribution and covering the same redshift range.

In order to take into account the cross-correlation between different LSS tracers, 
we extend the expression for the SNR \eqref{eqn:SNT} to:
\be
\left(\frac{S}{N}\right)^2_{\textrm{T,\,c$\times$g},\,\ell}
=\sum_{\rm X,Y=c,g}
C_\ell^\mathrm{TX}
\left[\text{Cov}_\ell^{-1}\right]_{\rm XY} \,,
C_\ell^\mathrm{TY}
\ee
where:
\be
\left[\text{Cov}_\ell\right]_{\rm XY} = 
\frac{C_\ell^\mathrm{TX}C_\ell^\mathrm{TY}+\bar{C}_\ell^\mathrm{TT}\bar{C}_\ell^\mathrm{XY}}
{\left(2\ell+1\right)\sqrt{f_\textrm{sky}^\textrm{X}f_\textrm{sky}^\textrm{Y}}} \,.
\ee
In this case the SNR by adding the information from Eu\_gal\_ph becomes 3.7 
for both Eu\_cl\_opt and Eu\_cl\_con. There is no actual improvement compared to 
using Eu\_gal\_ph alone but nevertheless it offer the possibility of a more robust 
ISW detection when cosmological parameters are not keep fixed.

The combinatin of the two Euclid cluster surveys with Eu\_gal\_sp brings a little improvement 
of +0.2 in the SNR.

Finally, we find that for eROSITA the combination of cluster survey with the galaxy survey 
from Euclid will be more promising in order to raise the SNR thanks to the low-z coverage from eROSITA. 
We find a SNR of 3.2/4.1 by combining and cross-corralating eR\_cl with Eu\_gal\_sp and Eu\_gal\_ph.

Also in this case the addition of the CMB polarization information increases the SNR by a further 
18\% for all the combinations.

\section{Cosmological forecasts with CMB-cluster cross-correlation}
\label{sec:six}
In this section we use the Fisher matrix technique \citep{Tegmark:1996bz} to compute the joint CMB 
and cross-correlation forecast constraints on few extensions of the $\Lambda$CDM cosmological 
concordance model.

\subsection{Methodology}
We use the Fisher matrix derived from the Gaussian approximation of the full likelihood ${\cal L}$. 
By using the Gaussian hypothesis for the data ${\bf d}$, the likelihood function of the cosmological 
parameters $\theta$ reads as:
\be
\label{enq:likelihood}
\mathcal{L}\left({\bf \theta}|{\bf d}\right) \propto \frac{1}{\sqrt{|{\bf C}\left({\bf \theta}\right)|}}
\text{exp}\left\{-\frac{1}{2}\left({\bf d}^\textrm{obs}\right)^\dagger \left[{\bf C}\left({\bf \theta}\right)\right]^{-1}{\bf d}^\textrm{obs}\right\} \,,
\ee
where:
\be
{\bf C} = 
\left( {\begin{array}{cccc}
C_\ell^\textrm{TT} & C_\ell^\textrm{TE} & C_\ell^\textrm{TX} \\
C_\ell^\textrm{TE} & C_\ell^\textrm{EE} & C_\ell^\textrm{EX} \\
C_\ell^\textrm{TX} & C_\ell^\textrm{EX} & C_\ell^\textrm{XX} \\
\end{array} } \right) \,,
\ee
is the theoretical covariance matrix of the modeled data 
${\bf d}=\left\{a_{\ell m}^\mathrm{T},a_{\ell m}^\mathrm{E},a_{\ell m}^\mathrm{X}\right\}$ 
where $X=\{\mathrm{c,g}\}$ and $|{\bf C}\left({\bf \theta}\right)|$ its determinant.
From Eq.~\eqref{enq:likelihood}, we write the Fisher matrix in compact form as:
\begin{align}
\label{eqn:fisher}
{\cal F}_{\alpha\beta} &= \langle \frac{\partial^2\log {\cal L}}{\partial\theta_\alpha\partial\theta_\beta} \rangle \,,\notag\\
&= \frac{1}{2} \text{tr} \left[\frac{\partial {\bf C}}{\partial\theta_\alpha}{\bf C}^{-1}\frac{\partial {\bf C}}{\partial\theta_\alpha}{\bf C}^{-1}\right]\,, \notag\\
&= \sum_{\ell}\frac{\left(2\ell+1\right)f_\textrm{sky}^\textrm{X}}{2}
\frac{\partial C_\ell^\textrm{AB}}{\partial \theta_\alpha} 
\text{Cov}^{-1}\left(C_\ell^\textrm{AB},C_\ell^\textrm{CD}\right) \frac{\partial C_\ell^\textrm{CD}}{\partial \theta_\beta} \,,
\end{align}
where $\bar{C}_\ell^{\rm AB} = C^{\rm AB}_\ell + \delta_\textrm{AB}N^{\rm A}_\ell$ is the sum of 
the signal and the noise, with ${\cal N}^{\rm TE}_\ell,{\cal N}^{\rm TX}_\ell,{\cal N}^{\rm EX}_\ell=0$.
For the temperature and polarization angular power spectra, here 
${\cal N}^{\rm T,E}_\ell=\sigma_{\rm T,E}\, b^{-2}_\ell$ 
is the isotropic noise deconvolved with the instrument beam, $b_\ell^2$ is the beam window function, 
assumed Gaussian, with $b_\ell = e^{-\ell (\ell + 1) \theta_{\rm FWHM}^2/16\ln 2}$; $\theta_{\rm FWHM}$ 
is the full width half maximum (FWHM) of the beam in radians; $\sigma_{\rm TT}$ and $\sigma_{\rm EE}$ 
are the square of the detector noise level on a steradian patch for temperature and polarization, 
respectively. See the appendix~\ref{sec:appendix} for the full structure of the covariance matrix 
Cov in Eq.~\eqref{eqn:fisher}.

As representative specifications for CMB we consider the $Planck$ 143 GHz channel full mission 
sensitivity and angular resolution as given in \cite{Adam:2015rua} and used in \cite{Ballardini:2016hpi}. 
These correspond to $\theta_{\rm FWHM}=7.3$ arcmin, $\sigma_{\rm T}=33$ arcmin$\,\mu$K 
and $\sigma_{\rm E}=70.2$ arcmin$\,\mu$K. 
For cluster and galaxy specifications we use those listed in Tab.~\ref{tab:Spec}.

We consider as set of cosmological parameters in our forecast analysis the six standard 
$\Lambda$CDM parameters $\omega_\textrm{b}$, $\omega_\textrm{c}$, $H_0$, $\tau$, 
$n_\textrm{s}$, and $\log\left(10^{10}\ A_\textrm{s}\right)$, plus some extra parameters described 
in the following subsection. We also consider the bias as a nuisance parameter marginalizing over it.

In the case of clusters, in addition to the variation on all the cosmological parameters we consider 
the dependence on four nuisance parameters $B_\textrm{M,\,0},\alpha,\sigma^2_{\log\,M,0},\,\beta$ 
introduced in Sec.~\ref{sec:nuisances}, as sources of uncertainties on the cluster selection function, 
as in \cite{Sartoris:2015aga}, by marginalizing over them.

\subsection{Results}
In this paper we restrict ourselves to forecast the joint CMB and ISW-cluster cross-correlation 
constraints on few relevant cosmological models as in \cite{2008A&A...485..395D}. With this setting 
the impact of ISW-cluster cross-correlation can be easily evaluated, whereas it would be hardly 
visible if we were considering the full combination of CMB and cluster cosmological information. 
As for the SNR in the previous section, we compare these results with those obtained by the joint 
CMB and ISW-galaxy cross-correlation expected from the Euclid spectroscopic and photometric surveys.

\begin{figure}
\centering
\hspace*{0.3cm}    
\includegraphics[width=7.8cm]{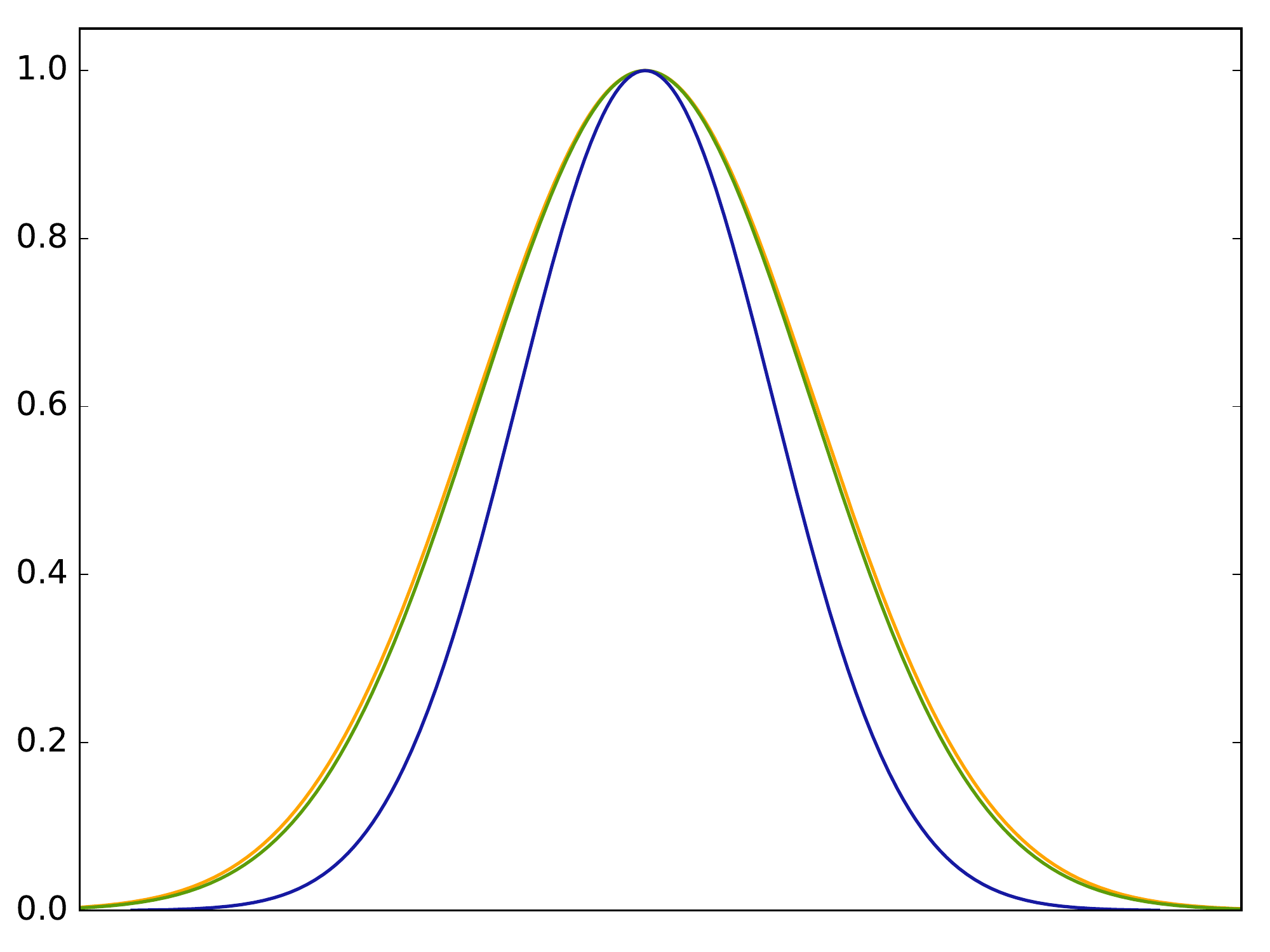}\\
\vspace*{-0.2cm}
\includegraphics[width=\columnwidth]{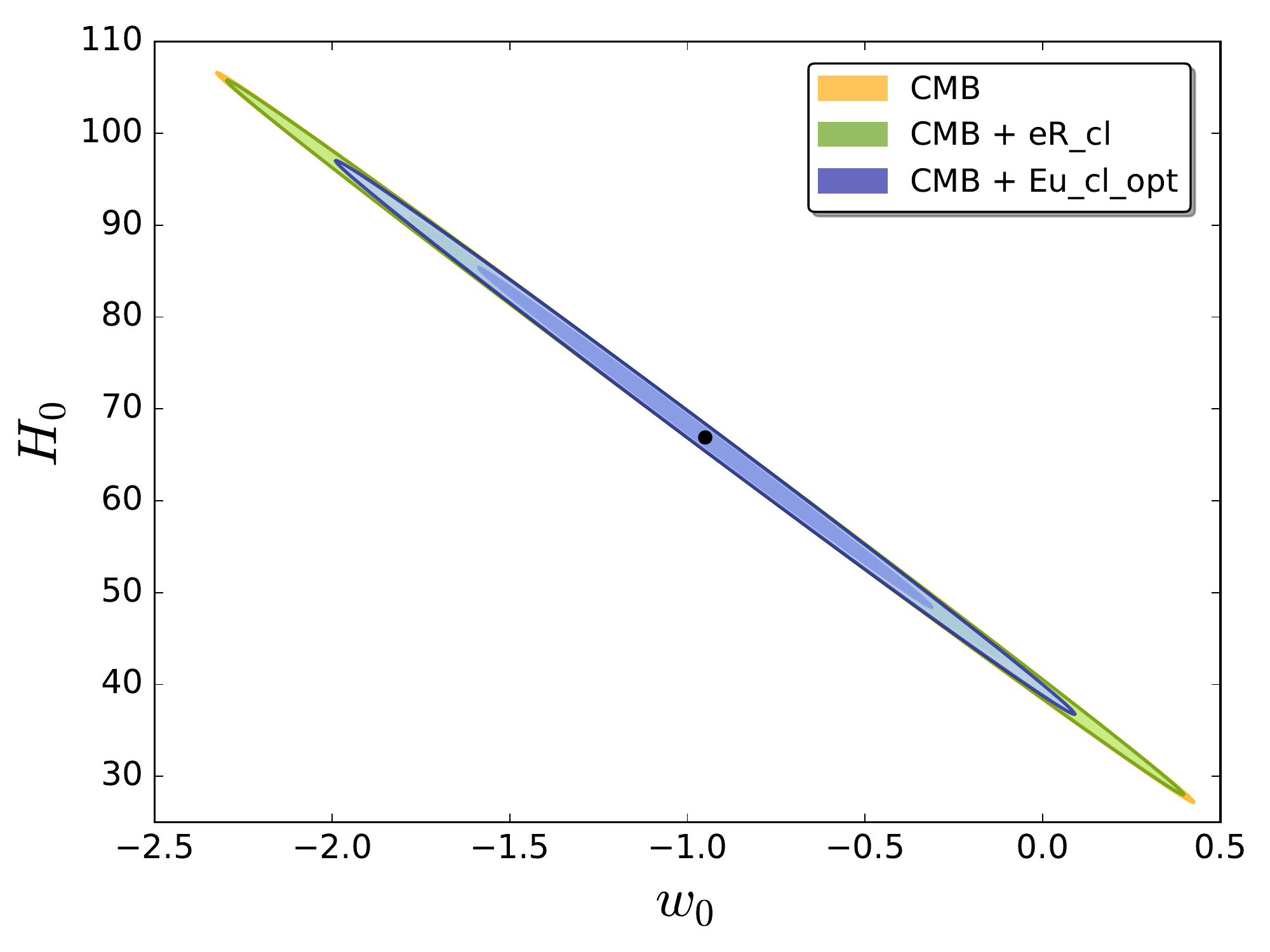}
\caption{In the bottom panel, we show the joint contraints on $\left(w_0,H_0\right)$ at the 68\,\% 
and 95\,\% CL. 
We show forecasts for the combination of the CMB and the CMB-LSS cross-correlation from eROSITA (green) 
and from the optimistic Euclid photometric cluster survey (blue). We show also the results for the 
CMB alone (yellow). 
In the top panel, we show the corresponding posterior distributions for $w_0$.}
\label{fig:w0h0}
\end{figure}

We first forecast the constraints on the redshift-independent parameter of state of dark energy 
$w_0$. Fig.~\ref{fig:w0h0} shows how the CMB-cluster cross-correlation improves on the CMB results 
alone. Our results show that CMB-cluster cross-correlation could lead to constraints similar to the 
one expected from the CMB-galaxy one, improving the constraints on $w_0$ obtained from CMB data alone 
by around 20-30\,\%.

\begin{figure}
\centering
\includegraphics[width=\columnwidth]{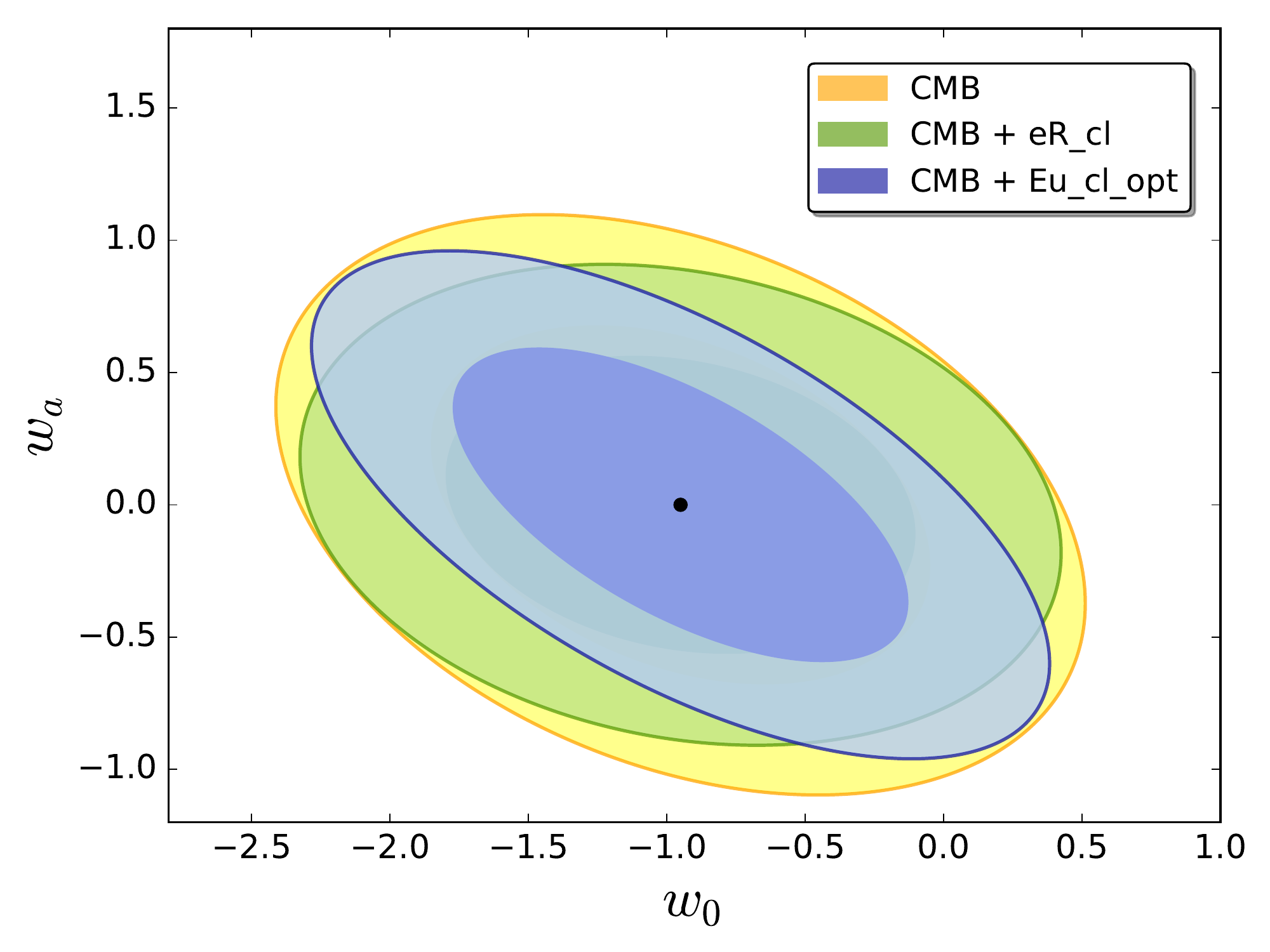}
\caption{Joint contraints on $\left(w_0,w_a\right)$ at the 68\,\% and 95\,\% CL. We show forecasts 
for the combination of the CMB and the CMB-LSS cross-correlation from eROSITA (green) and from the 
optimistic Euclid photometric cluster survey (blue). We show also the results for the CMB alone (yellow).}
\label{fig:w0wa}
\end{figure}

As second cosmological scenario we show the constraints for a cosmology in which the parameter 
of state of dark energy is allowed to be redshift-dependent \citep{Chevallier:2000qy,Linder:2003ec}:
\be
w(z) = w_0 + \frac{z}{1+z} w_a \,,
\ee
where $w_0$ is the present value of the equation of state and $w_a$ is its first derivative 
with respect to the scale parameter $a(t)$. Also in this case, we see from Fig.~\ref{fig:w0wa} 
that the addition of CMB-cluster cross-correlations leads to a significant improvement for the 
dark energy parameters, with a similar performance between eROSITA and Euclid.
The improvement with respect to CMB-only constraints for the $w_0-w_\mathrm{a}$ CDM cosmological 
model corresponds to: 6\,\% on $H_0$, 4\,\% over $n_\textrm{s}$, 5\,\% on $w_0$, and 20\,\% on $w_a$. 
The Figure of Merit (FoM) for the dark energy parameters \citep{Albrecht:2009ct} improves by a factor 
1.5 when the ISW-cluster cross-correlation is included.
In Fig.~\ref{fig:w0wa}, the difference between the orientation of the contour of eROSITA and the one of Euclid 
in the $w_0-w_\mathrm{a}$ parameters space is not only due to the different redshift range probed by the two surveys but it is also due to the larger sky coverage from eROSITA which helps in breaking the degeneracy between the two parameters.

We consider a cosmological model in which there is a suppression on large 
scales of the primodial curvature power spectrum \citep{Contaldi:2003zv}:
\begin{equation}
\mathcal{P_R}(k) = \mathcal{P}_{{\cal R},\,0}(k) \left\{ 1 - \exp \left[- \left( \frac{k}{k_{\rm c}}
\right)^{\lambda_{\rm c}} \right] \right\} \,,
\label{eqn:cutoff}
\end{equation}
where $k_{\rm c}$, is the relevant scale of deviation from a power-law spectrum and $\lambda_{\rm c}$ 
adjusts the stiffness of the suppression.
For this model the addition of the CMB-cluster cross-correlation leads to a hardly visibile 
improvement in the constraints on $(\lambda_{\rm c}, k_{\rm c})$ with respect to the CMB constraints, 
since, even if on large scales, the suppression induces a change in the SW 
contribution rather than on the ISW one. Nevertheless, the ISW-LSS cross-correlation is useful to break 
degeneracies in the physical effects at play on the largest scales.

%\begin{figure}
%\centering
%\includegraphics[width=\columnwidth]{XCMB_cutoff_v3.pdf}
%\caption{Joint contraints on $\left(\log\left(k_\textrm{c}\textrm{ Mpc}\right),\lambda_\textrm{c}\right)$ 
%at the 68\,\% and 95\, \% CL. We show forecasts for the combination of the CMB and the CMB-LSS 
%cross-correlation from the optimistic Euclid photometric cluster survey (blue). We show also the 
%results for the CMB alone (yellow).}
%\label{fig:cutoff}
%\end{figure}
Note that in our calculations we have taken into account the dependence of both the halo mass and 
bias fuctions on the underlying cosmology through the variance of the fractional density fluctuation 
$\sigma^2(M,z)$. We consider this dependence when varying the angular power spectra $C_\ell^\textrm{cc}$ 
and $C_\ell^\textrm{Tc}$ around the fiducial cosmology in the calculation of the Fisher matrix 
\eqref{eqn:fisher}. The effect of neglecting this dependence would lead to an increase in the 
uncertainties on the cosmological parameters of about 10\,\% for the most significant parameters 
such as $w_0$, and $w_\textrm{a}$. 

\begin{figure}
\centering
\includegraphics[width=\columnwidth]{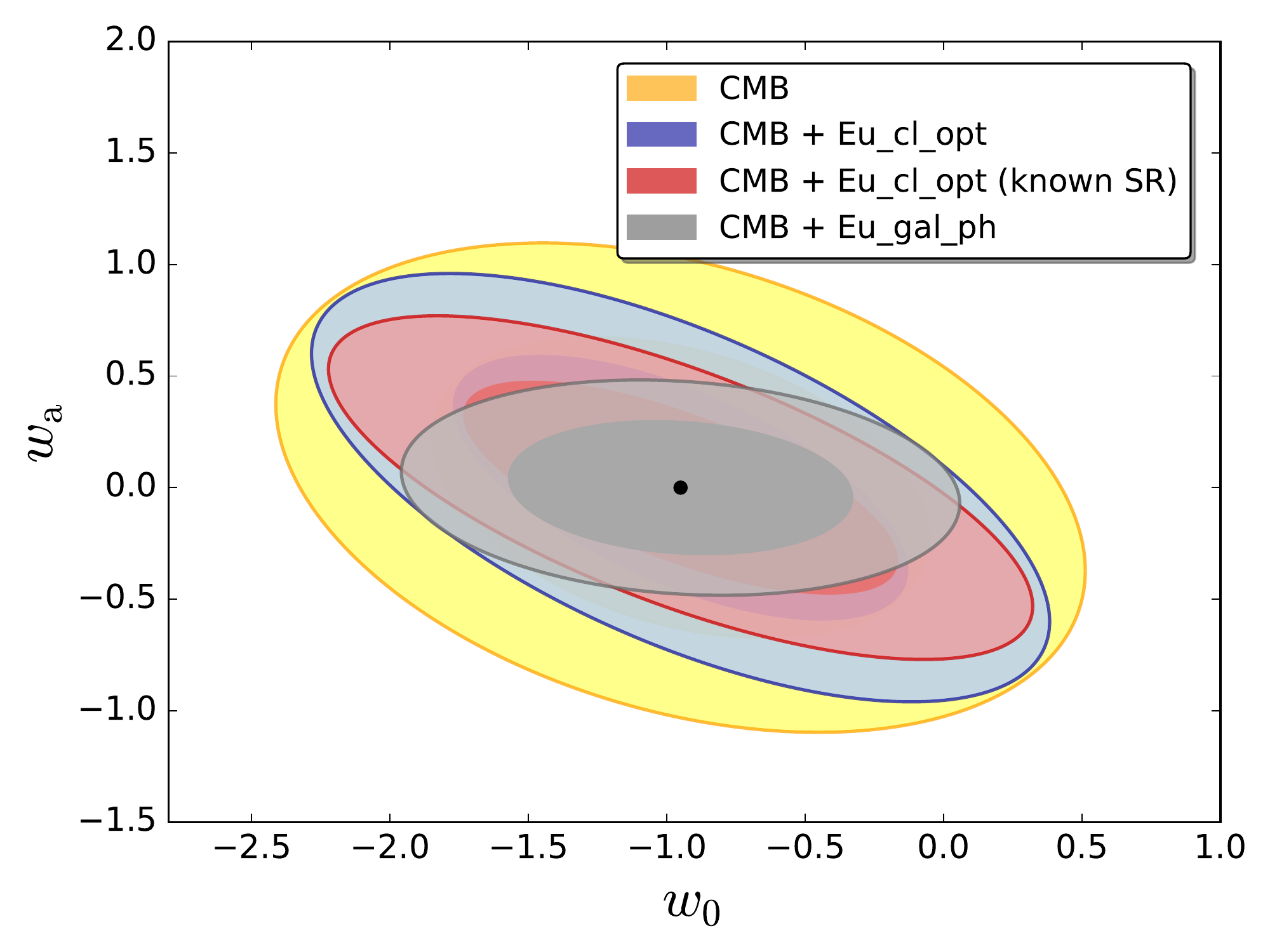}
\caption{Joint contraints on $\left(w_0,w_\textrm{a}\right)$ at the 68\,\% and 95\,\% CL.
We show forecasts for the combination of the CMB and the CMB-LSS cross-correlation from the 
optimistic Euclid photometric cluster survey with (red) and without (blue) assuming a perfect 
knowledge of the scaling relation, and from the photometric Euclid galaxy survey (gray). We also show 
the results for the CMB alone for comparison (yellow).}
\label{fig:Galw0wa}
\end{figure}

As a semi-idealized case, we also consider the constraints on the $w_0-w_\mathrm{a}$ plane by 
assuming perfect knowledge of the scaling relation between the true and the observed cluster mass, 
i.e. by fixing all the corresponding nuisance parameters: this semi-idealized case would cast the 
cluster case at the same footing of the galaxy one, in which no nuisance parameter is considered. 
In Fig.~\ref{fig:Galw0wa} we show the comparison of the constraints in the $w_0$-$w_\textrm{a}$ 
parameters space between the CMB-cluster cross-correlation from Eu\_cl\_opt, with and without 
perfect knowledge of the scaling relation, and the CMB-galaxy cross-correlation from Eu\_gal\_ph. 
Assuming a perfect knowledge of the nuisance parameters connected to the cluster selection function 
leads to an improvement in the constraints obtainable with the help the CMB-cluster cross-correlation. 
In particular, we found an improvement of 10\,\% for $w_0$, of 20\,\% and 40\,\% for $w_0$ and $w_a$, 
of 5\,\% for $k_c$ and $\lambda_c$.

\section{Conclusions}
\label{sec:concl}
Cosmology with clusters is a rapidly evolving field, thanks to improvements in observations. 
Current catalogues are already sufficient to obtain an accurate measurement of the baryon 
acoustic oscillation with galaxy clusters as from the Sloan Digital Sky Survey leading to a 
distance-redshift relation in full agreement with $Planck$ observations and with uncertainties 
similar to the one obtained by galaxy surveys \citep{Veropalumbo:2013cua,Veropalumbo:2015dpi}. 
In the perspective of future surveys in several wavelengths which will provide catalogues with 
a higher number of clusters over a wider redshift range, we have studied the capabilities to 
detect the ISW effect by a CMB-cluster cross-correlation. Note that this ISW-cluster 
cross-correlation is not internal to CMB as the previously studied ISW-tSZ cross-correlation 
\citep{Taburet:2010hb}.
In this perspective, it will be essential to control the SZ cluster residuals in CMB maps that 
could bias the cross-correlation measurements between clusters and CMB \citep{Chen:2018drw}.

As two representative cases, we have considered the specifications of the cluster catalogues as 
expected from two coming space missions, i.e. eROSITA \citep{2010SPIE.7732E..0UP} and Euclid 
\citep{Laureijs:2011gra}. We have found that the CMB-cluster cross-correlation could be used for 
a statistically significant detection of the ISW effect.
As two general remarks, we have found that: (1) the signal-to-noise of the CMB-cluster cross-correlation 
benefits of the large cluster bias which balances the larger shot noise in cluster catalogues 
compared to galaxy surveys; (2) the dependence of the cluster bias on cosmology needs to be taken 
into account to extract the full information from the cross-correlation with CMB.
Whereas for eROSITA the cross-correlation with CMB would be an added value for the legacy of its 
cluster catalogue, for Euclid it would constitute a sort of coarse graining of the CMB galaxy 
cross-correlation and a key verification of the structure formation process. 
Nevertheless, the CMB-cluster would be another valuable cross-correlation within the Euclid 
probes and would add useful information on the bias for the cluster catalogue. The cross-correlation 
studied here would be of interest for other surveys as well, as LSST \citep{Abell:2009aa} and SKA 
\citep{Maartens:2015mra}.

\section*{Acknowledgements}
We thank Nabila Aghanim and Carlo Baccigalupi for useful comments to the paper.
We acknowledge financial contribution from the agreement ASI n.I/023/12/0 "Attivit\`a relative 
alla fase B2/C per la missione Euclid". The support by the "ASI/INAF Agreement 2014-024-R.0 for 
the Planck LFI Activity of Phase E2" is also acknowledged.
MB, DP, FF and LM, acknowledge the support from the grant MIUR PRIN 2015
"Cosmology and Fundamental Physics: illuminating the Dark Universe with Euclid".
MB was supported by the South African Radio Astronomy Observatory, which is a facility 
of the National Research Foundation, an agency of the Department of Science and Technology 
and the Claude Leon Foundation. BS acknowledges financial support from 
the University of Trieste through the program "Finanziamento di Ateneo per progetti di ricerca 
scientifica - FRA 2015", a grant from "Consorzio per la Fisica - Trieste" and from the PRIN 2015W7KAWC 
project, funded by the Italian Minister for University and Research.

%%%%%%%%%%%%%%%%%%%%%%%%%%%%%%%%%%%%%%%%%%%%%%%%%%



\bibliographystyle{mnras}
\bibliography{Biblio} % if your bibtex file is called example.bib

%\begin{thebibliography}{99}
%\end{thebibliography}

%%%%%%%%%%%%%%%%%%%%%%%%%%%%%%%%%%%%%%%%%%%%%%%%%%

%%%%%%%%%%%%%%%%% APPENDICES %%%%%%%%%%%%%%%%%%%%%

\appendix

\section{The covariance in the Fisher approach}
\label{sec:appendix}
Since we are interested in the effect coming from 
the cross-correlation between the CMB and LSS surveys alone, we do not consider information 
coming from the LSS tracer angular power spectrum, i.e. we use $C_\ell^\textrm{XX}$ just for the 
covariance.
The non-vanishing elements of the symmetric angular power spectrum covariance matrix $\text{Cov}_\ell$ 
at the $\ell^\text{th}$ multipole are:
\be
\text{Cov}\left(C_\ell^\textrm{AB},C_\ell^\textrm{CD}\right) = \bar{C}_\ell^\textrm{AC}\bar{C}_\ell^\textrm{BD} + \bar{C}_\ell^\textrm{AD}\bar{C}_\ell^\textrm{BC} \,.
\ee

The exact structure of $\text{Cov}^{-1}_\ell$ is more useful compared to its inverse since it allows 
to derive all the reduced cases for its inverse when some cross-correlation terms go to zero:
\begin{align}
&\text{Cov}^{-1}_\ell = \frac{1}{\left(|{\bf C}\left({\bf \theta}\right)|\right)^2} \notag\\
&\left( {\begin{array}{cccccc}
\Sigma_\textrm{TTTT} & \Sigma_\textrm{TTEE} & \Sigma_\textrm{TTTE} & \Sigma_\textrm{TTXX} & \Sigma_\textrm{TTTX} & \Sigma_\textrm{TTEX}\\
\Sigma_\textrm{TTEE} & \Sigma_\textrm{EEEE} & \Sigma_\textrm{EETE} & \Sigma_\textrm{EEXX} & \Sigma_\textrm{EETX} & \Sigma_\textrm{EEEX}\\
\Sigma_\textrm{TTTE} & \Sigma_\textrm{EETE} & \Sigma_\textrm{TETE} & \Sigma_\textrm{TEXX} & \Sigma_\textrm{TETX} & \Sigma_\textrm{TEEX}\\
\Sigma_\textrm{TTXX} & \Sigma_\textrm{EEXX} & \Sigma_\textrm{TEXX} & \Sigma_\textrm{XXXX} & \Sigma_\textrm{XXTX} & \Sigma_\textrm{XXEX}\\
\Sigma_\textrm{TTTX} & \Sigma_\textrm{EETX} & \Sigma_\textrm{TETX} & \Sigma_\textrm{XXTX} & \Sigma_\textrm{TXTX} & \Sigma_\textrm{TXEX}\\
\Sigma_\textrm{TTEX} & \Sigma_\textrm{EEEX} & \Sigma_\textrm{TEEX} & \Sigma_\textrm{XXEX} & \Sigma_\textrm{TXEX} & \Sigma_\textrm{EXEX}\\
\end{array} } \right) \,,
\end{align}
where the auto-correlation are:
\begin{align}
\Sigma_\textrm{TTTT}
&= \left[
\left(\bar{C}_\ell^{\rm EX}\right)^2-\bar{C}_\ell^{\rm EE}\bar{C}_\ell^{\rm XX}\right]^2 \,,\\
\Sigma_\textrm{EEEE}
&= \left[
\left(\bar{C}_\ell^{\rm TX}\right)^2-\bar{C}_\ell^{\rm TT}\bar{C}_\ell^{\rm XX}\right]^2 \,,\\
\Sigma_\textrm{TETE}
&= 2\left\{2\left(\bar{C}_\ell^{\rm TX}\right)^2\left(\bar{C}_\ell^{\rm EX}\right)^2\right. \notag\\
&\qquad\left.
+\left(\bar{C}_\ell^{\rm XX}\right)^2\left[\left(\bar{C}_\ell^{\rm TE}\right)^2-
\bar{C}_\ell^{\rm TT}\bar{C}_\ell^{\rm EE}\right] \right.\notag\\
&\qquad\left.-\bar{C}_\ell^{\rm XX}\left[
2\bar{C}_\ell^{\rm TE}\bar{C}_\ell^{\rm TX}\bar{C}_\ell^{\rm EX}
+\bar{C}_\ell^{\rm TT}\left(\bar{C}_\ell^{\rm EX}\right)^2 \right.\right.\notag\\
&\qquad\left.\left.+\bar{C}_\ell^{\rm EE}\left(\bar{C}_\ell^{\rm TX}\right)^2\right]\right\} \,,\\
\Sigma_\textrm{XXXX}
&= \left[
\left(\bar{C}_\ell^{\rm TE}\right)^2-\bar{C}_\ell^{\rm TT}\bar{C}_\ell^{\rm EE}\right]^2 \,,\\
\Sigma_\textrm{TXTX}
&= 2\left\{
-2\bar{C}_\ell^{\rm EE}\bar{C}_\ell^{\rm TE}\bar{C}_\ell^{\rm TX}\bar{C}_\ell^{\rm EX} \right. \notag\\
&\qquad\left.+\left(\bar{C}_\ell^{\rm EX}\right)^2\left[2\left(\bar{C}_\ell^{\rm TE}\right)^2 -
\bar{C}_\ell^{\rm TT}\bar{C}_\ell^{\rm EE}\right] \right.\notag\\
&\qquad\left.+\bar{C}_\ell^{\rm EE}\left[
\bar{C}_\ell^{\rm TT}\bar{C}_\ell^{\rm EE}\bar{C}_\ell^{\rm XX}
+\bar{C}_\ell^{\rm EE}\left(\bar{C}_\ell^{\rm TX}\right)^2 \right.\right.\notag\\
&\qquad\left.\left.-\bar{C}_\ell^{\rm XX}\left(\bar{C}_\ell^{\rm TE}\right)^2\right]\right\} \,,\\
\Sigma_\textrm{EXEX}
&= 
2\left\{
2\left(\bar{C}_\ell^{\rm TE}\right)^2\left(\bar{C}_\ell^{\rm TX}\right)^2 
-\bar{C}_\ell^{\rm XX}\left(\bar{C}_\ell^{\rm TE}\right)^2\right.\notag\\
&\qquad\left.+\left(\bar{C}_\ell^{\rm TT}\right)^2\left[\left(\bar{C}_\ell^{\rm EX}\right)^2+
\bar{C}_\ell^{\rm EE}\bar{C}_\ell^{\rm XX}\right] \right.\notag\\
&\qquad\left.-\bar{C}_\ell^{\rm TX}\left(
2\bar{C}_\ell^{\rm TE}\bar{C}_\ell^{\rm EX}
+\bar{C}_\ell^{\rm EE}\bar{C}_\ell^{\rm TX}\right)\right\} \,,
\end{align}
and the cross-correlation are:
\begin{align}
\Sigma_\textrm{TTEE}
&= \left[
\bar{C}_\ell^{\rm XX}\bar{C}_\ell^{\rm TE}-\bar{C}_\ell^{\rm EX}\bar{C}_\ell^{\rm TX}\right]^2 \,,\\
\Sigma_\textrm{TTTE}
&= 2\left[
\left(\bar{C}_\ell^{\rm EX}\right)^2-\bar{C}_\ell^{\rm EE}\bar{C}_\ell^{\rm XX}\right]\left(
\bar{C}_\ell^{\rm TE}\bar{C}_\ell^{\rm XX}-\bar{C}_\ell^{\rm TX}\bar{C}_\ell^{\rm EX}\right) \,,\\
\Sigma_\textrm{TTXX}
&= \left(
\bar{C}_\ell^{\rm TE}\bar{C}_\ell^{\rm EX}-\bar{C}_\ell^{\rm EE}\bar{C}_\ell^{\rm TX}\right)^2 \,,\\
\Sigma_\textrm{TTTX}
&= 2\left[
\left(\bar{C}_\ell^{\rm EX}\right)^2-\bar{C}_\ell^{\rm EE}\bar{C}_\ell^{\rm XX}\right]\left(
\bar{C}_\ell^{\rm EE}\bar{C}_\ell^{\rm TX}-\bar{C}_\ell^{\rm TE}\bar{C}_\ell^{\rm EX}\right) \,,\\
\Sigma_\textrm{TTEX}
&= 2\left(
\bar{C}_\ell^{\rm TE}\bar{C}_\ell^{\rm EX}-\bar{C}_\ell^{\rm EE}\bar{C}_\ell^{\rm TX}\right)
\left(
\bar{C}_\ell^{\rm TX}\bar{C}_\ell^{\rm EX}-\bar{C}_\ell^{\rm TE}\bar{C}_\ell^{\rm XX}\right) \,,
\end{align}
\begin{align}
\Sigma_\textrm{EETE}
&= 2\left[
\left(-\bar{C}_\ell^{\rm TX}\right)^2+\bar{C}_\ell^{\rm TT}\bar{C}_\ell^{\rm XX}\right]\left(
\bar{C}_\ell^{\rm TX}\bar{C}_\ell^{\rm EX}-\bar{C}_\ell^{\rm TE}\bar{C}_\ell^{\rm XX}\right) \,,\\
\Sigma_\textrm{EEXX}
&= \left(
\bar{C}_\ell^{\rm TE}\bar{C}_\ell^{\rm TX}-\bar{C}_\ell^{\rm TT}\bar{C}_\ell^{\rm EX}\right)^2 \,,\\
\Sigma_\textrm{EETX}
&= 2\left(
\bar{C}_\ell^{\rm TE}\bar{C}_\ell^{\rm XX}-\bar{C}_\ell^{\rm TX}\bar{C}_\ell^{\rm EX}\right)
\left(\bar{C}_\ell^{\rm TT}\bar{C}_\ell^{\rm EX}-\bar{C}_\ell^{\rm TE}\bar{C}_\ell^{\rm TX}\right) \,,\\
\Sigma_\textrm{EEEX}
&= 2\left[
\left(\bar{C}_\ell^{\rm TX}\right)^2-\bar{C}_\ell^{\rm TT}\bar{C}_\ell^{\rm XX}\right]\left(
\bar{C}_\ell^{\rm TT}\bar{C}_\ell^{\rm EX}-\bar{C}_\ell^{\rm TE}\bar{C}_\ell^{\rm TX}\right) \,,
\end{align}
\begin{align}
\Sigma_\textrm{TEXX}
&= 2\left(
\bar{C}_\ell^{\rm EE}\bar{C}_\ell^{\rm TX}-\bar{C}_\ell^{\rm TE}\bar{C}_\ell^{\rm EX}\right)
\left(\bar{C}_\ell^{\rm TT}\bar{C}_\ell^{\rm EX}-\bar{C}_\ell^{\rm TE}\bar{C}_\ell^{\rm TX}\right) \,,\\
\Sigma_\textrm{TETX}
&= -2\left\{
-2\bar{C}_\ell^{\rm EE}\bar{C}_\ell^{\rm TE}\bar{C}_\ell^{\rm XX}\bar{C}_\ell^{\rm TX}
-\bar{C}_\ell^{\rm TT}\left(\bar{C}_\ell^{\rm EX}\right)^3 \right.\notag\\
&\qquad\left.+\bar{C}_\ell^{\rm EX}\left[\left(\bar{C}_\ell^{\rm TX}\right)^2\bar{C}_\ell^{\rm EE} \right.\right.\notag\\
&\qquad\left.\left.
+\bar{C}_\ell^{\rm XX}\left[\left(\bar{C}_\ell^{\rm TE}\right)^2+\bar{C}_\ell^{\rm TT}\bar{C}_\ell^{\rm EE}\right]\right]\right\} \,,\\
\Sigma_\textrm{TEEX}
&= -2\left\{
\left(\bar{C}_\ell^{\rm EX}\right)^2\bar{C}_\ell^{\rm TT}\bar{C}_\ell^{\rm TX}
-\bar{C}_\ell^{\rm EE}\left(\bar{C}_\ell^{\rm TX}\right)^3 \right.\notag\\
&\qquad\left.
+\bar{C}_\ell^{\rm XX}\left[\left(\bar{C}_\ell^{\rm TE}\right)^2\bar{C}_\ell^{\rm TX}
-2\bar{C}_\ell^{\rm TT}\bar{C}_\ell^{\rm TE}\bar{C}_\ell^{\rm EX} \right.\right.\notag\\
&\qquad\left.\left.+\bar{C}_\ell^{\rm TT}\bar{C}_\ell^{\rm EE}\bar{C}_\ell^{\rm TX}\right]\right\} \,,
\end{align}
\begin{align}
\Sigma_\textrm{XXTX}
&= 2\left[
\left(\bar{C}_\ell^{\rm TE}\right)^2-\bar{C}_\ell^{\rm EE}\bar{C}_\ell^{\rm TT}\right]\left(
\bar{C}_\ell^{\rm EE}\bar{C}_\ell^{\rm TX}-\bar{C}_\ell^{\rm TE}\bar{C}_\ell^{\rm EX}\right) \,,\\
\Sigma_\textrm{XXEX}
&= 2\left[
\left(\bar{C}_\ell^{\rm TE}\right)^2-\bar{C}_\ell^{\rm EE}\bar{C}_\ell^{\rm TT}\right]\left(
\bar{C}_\ell^{\rm TT}\bar{C}_\ell^{\rm EX}-\bar{C}_\ell^{\rm TX}\bar{C}_\ell^{\rm TE}\right) \,,\\
\Sigma_\textrm{TXEX}
&= -2\left\{
\left(\bar{C}_\ell^{\rm EX}\right)^2\bar{C}_\ell^{\rm TT}\bar{C}_\ell^{\rm TE} \right.\notag\\
&\qquad\left.
+\bar{C}_\ell^{\rm EE}\bar{C}_\ell^{\rm TX}\left(\bar{C}_\ell^{\rm TE}\bar{C}_\ell^{\rm TX}
-2\bar{C}_\ell^{\rm TT}\bar{C}_\ell^{\rm EX}\right) \right.\notag\\
&\qquad\left.
+\bar{C}_\ell^{\rm XX}\left[\bar{C}_\ell^{\rm TT}\bar{C}_\ell^{\rm EE}\bar{C}_\ell^{\rm TE}
-\left(\bar{C}_\ell^{\rm TE}\right)^3\right]\right\} \,.
\end{align}


%\input app_section_three.tex

%%%%%%%%%%%%%%%%%%%%%%%%%%%%%%%%%%%%%%%%%%%%%%%%%%

% Don't change these lines
\bsp	% typesetting comment
\label{lastpage}
\end{document}